\documentclass[twocolumn,groupedaddress,superscriptaddress,amsfonts, amssymb, amsmath, caption]{revtex4-1}

\usepackage{mathtools}

\usepackage[noplusnominus,noequal,noasterisk,nospecials,nolessnomore,italic]{mathastext}
\usepackage[stretch=17,shrink=17,step=1,tracking,kerning,final]{microtype}  % enable character protrusion and font expansion

\usepackage{cmap}  % makes ligatures etc. work with PDF viewers (i.e. makes the PDF copyable and searchable)
\everymath{\displaystyle} %% Displays math a full height, even when in-line with text
\DeclareSymbolFont{UPM}{U}{eur}{m}{n}
\DeclareMathSymbol{\partial}{0}{UPM}{"40}
\usepackage{xfrac} %% \sfrac command for side fractions

\usepackage[version=3]{mhchem}  %% to typeset chemical formulae easily (using \ce{})

\usepackage[range-phrase=-, range-units=single]{siunitx} %% SI units
\DeclareSIUnit\molar{\mole\per\cubic\deci\metre}
\DeclareSIUnit\Molar{\textsc{m}}
\DeclareSIUnit\kT{$k_{B}T$}
\DeclareSIUnit\bp{}
\usepackage{xr} % No hyper-ref
\usepackage{hyperref}
\usepackage{cleveref} %% Clever reference: http://tex.stackexchange.com/questions/83037/difference-between-ref-varioref-and-cleveref-decision-for-a-thesis
\crefname{equation}{eq.}{eqs.}
\Crefname{equation}{Eq.}{Eqs.}
\usepackage{placeins}

\usepackage[UKenglish]{babel} %% Controls headers and hyphanation patterns
\usepackage[utf8]{inputenc} % Uses Unicode instead of ASCII

\usepackage{paralist}
\setdefaultenum{(i)}{(a)}{(1)}{(A)}
\usepackage{afterpage} % Full page figures and tables
\usepackage{tabularx, booktabs}
\usepackage{longtable}
\newcolumntype{R}{>{\raggedleft\arraybackslash}X}
\newcommand{\ra}[1]{\renewcommand{\arraystretch}{#1}}

\newcommand{\etal}{{\textit{et al.~}}}
\newcommand{\etalnospace}{{\textit{et al.}}}
\newcommand{\etalnoperiod}{{\textit{et al}}}

\newcommand{\superscript}[1]{\ensuremath{^{\textrm{#1}}}}

\newcommand{\textcaption}[1]{\textrm{#1}} % Caption text

\def\yieldtoA(#1){-ln(#1/(1-#1))}%

\newcommand{\beginsupplement}{%
    \setcounter{table}{0}
    \renewcommand{\thetable}{S\arabic{table}}%
    \setcounter{figure}{0}
    \renewcommand{\thefigure}{S\arabic{figure}}%
    \setcounter{section}{0}
    \renewcommand{\thesection}{S\arabic{section}}%
    \setcounter{equation}{0}
    \renewcommand{\theequation}{S\arabic{equation}}%
    \setcounter{page}{1}
}

\begin{document}

\title{Coarse-grained modelling of strong DNA bending I: Thermodynamics and comparison to an experimental ``molecular vice''}

\author{Ryan M. Harrison}
\affiliation{Physical \& Theoretical Chemistry Laboratory, Department of Chemistry, University of Oxford, South Parks Road, Oxford OX1 3QZ, United Kingdom} 

\author{Flavio Romano}
\affiliation{Physical \& Theoretical Chemistry Laboratory, Department of Chemistry, University of Oxford, South Parks Road, Oxford OX1 3QZ, United Kingdom} 

\author{Thomas E. Ouldridge}
\affiliation{Department of Mathematics, Imperial College, 180 Queen's Road, London SW7 2AZ, United Kingdom}
\affiliation{Rudolf Peierls Centre for Theoretical Physics, Department of Physics, University of Oxford, 1 Keble Road, Oxford OX1 3NP, United Kingdom}

\author{Ard A. Louis}
\affiliation{Rudolf Peierls Centre for Theoretical Physics, Department of Physics, University of Oxford, 1 Keble Road, Oxford OX1 3NP, United Kingdom}

\author{Jonathan P. K. Doye}
\affiliation{Physical \& Theoretical Chemistry Laboratory, Department of Chemistry, University of Oxford, South Parks Road, Oxford OX1 3QZ, United Kingdom}

\date{\today}

\begin{abstract}

DNA bending is biologically important for genome regulation and is relevant to
a range of nanotechnological systems. Recent results suggest that sharp bending
is much easier than implied by the widely-used worm-like chain model; many of
these studies, however, remain controversial. We use a coarse-grained model,
previously fitted to DNA's basic thermodynamic and mechanical properties, to
explore strongly bent systems. We find that as the end-to-end distance is
decreased sufficiently short duplexes undergo a transition to a state in which
the bending strain is localized at a flexible kink that involves disruption of
base-pairing and stacking.  This kinked state, which is not well-described by
the worm-like chain model, allows the duplex to more easily be sharply bent.
It is not completely flexible, however, due to constraints arising from the
connectivity of both DNA backbones.  We  also perform a detailed comparison to
recent experiments on a "molecular vice" that probes highly bent DNA.  Close
agreement between simulations and experiments strengthens the hypothesis that
localised bending via kinking occurs in the molecular vice and causes enhanced
flexibility of duplex DNA. Our calculations therefore suggests that the cost of
kinking implied by this experiment is consistent with the known thermodynamic
and mechanical properties of DNA.

\end{abstract}

\pacs{}

\maketitle

\section*{Introduction}

Strong bending of DNA plays a critical biological role in genome maintenance and regulation, most famously evident in DNA-protein complexes such as those formed in conjunction with the Lac repressor \cite{lewis_crystal_1996}, the TATA binding protein \cite{nikolov_crystal_1996} and the nucleosome \cite{richmond_structure_2003, widom_role_2001}. Experiments have confirmed that these large bending fluctuations are not an artefact of DNA-protein co-crystallization, and are observed in a variety of sensitive solution-phase assays \cite{haeusler_fret_2012, davis_tata_1999}. In the regime where DNA is not strongly bent, 
worm-like chain (WLC) models \cite{shimada_ring-closure_1984, thirumalai_statistical_1997, mazur_wormlike_2007, becker_radial_2010} 
that treat DNA as a semi-flexible rod are sufficient to describe its configurational behaviour \cite{shore_energetics_1983,Bustamante94,du_cyclization_2005}. It has been claimed, however, that strongly bent configurations can be reached much more easily than would be predicted by the WLC as parameterized to small fluctuations. In particular, deviations from WLC behaviour have been reported in experiments on cyclization \cite{cloutier_spontaneous_2004, vafabakhsh_extreme_2012, le_probing_2014}, multimerization \cite{podtelezhnikov_multimerization-cyclization_2000}, in DNA minicircles \cite{du_kinking_2008, demurtas_bending_2009}, in a stressed ring system \cite{qu_elastic_2010, qu_critical_2011, qu_complete_2011,kim15} and a molecular vice \cite{fields_euler_2013}. 

Much of the evidence for super-WLC flexibility, has, however, been contested.
For example, in DNA cyclization assays, cyclized molecules with $\sim 100$ base
pairs (bp) \cite{cloutier_spontaneous_2004, vafabakhsh_extreme_2012} were
reported to be orders of magnitude more likely than predicted by the WLC model
as normally parameterized. However, other authors have criticized the
interpretation of the experimental results and have suggested that non-WLC
behavior only begins at shorter lengths \cite{du_cyclization_2005,
vologodskii_bending_2013,vologodskii_strong_2013}. Similarly, apparent evidence of high flexibility of
DNA on short length scales obtained via AFM \cite{wiggins_high_2006} has since
been attributed to artefacts of the curve-fitting algorithms
\cite{mazur_dna_2014}.

The WLC model, which assumes a harmonic cost on the local degree of bending,
favours the distribution of bending strain throughout a DNA molecule. However,
above a certain degree of local bending the WLC model may break down because it
becomes favourable for DNA systems to localize strain within small regions,
allowing the remainder to relax and reducing the overall cost of adopting
highly bent configurations. A disruption of the B-DNA structure (including
broken base pairs and/or stacking interactions), or ``kink'', is a candidate
motif that would allow such enhanced flexibility and localization of bending
\cite{crick_kinky_1975, yan_localized_2004, du_kinking_2008,
mitchell_atomistic_2011, sivak_consequences_2011, vologodskii_strong_2013}. 

Nevertheless, although there is experimental evidence for kinking in strongly
bent configurations (reviewed for example in
reference \cite{vologodskii_strong_2013}), it is not clear whether all of the
proposed violations of WLC behaviour in the experiments listed above can be
explained in this way.  One reason for the lack of clarity is that a number of
fundamental questions about the thermodynamic cost and basic physics associated
with the formation of kinks remain open. 
It is therefore important to develop a sound theoretical underpinning of the
underlying physics. 
Doing so would enable a systematic comparison between distinct systems and help
establish whether experimental results are self-consistent and also consistent
with basic DNA thermodynamics. 

Here we use oxDNA
\cite{ouldridge_structural_2011,ouldridge_coarse-grained_2012,sulc_sequence-dependent_2012},
a coarse-grained model of DNA, to address these questions. OxDNA incorporates a
physical description of single- and double-stranded DNA, including the
thermodynamics of hybridization and basic mechanical properties such as
persistence length and torsional modulus \cite{ouldridge_structural_2011,
ouldridge_coarse-grained_2012}. As such, it is an ideal model for exploring
strong bending, highlighting general properties and providing evidence as to
whether the reported observations of enhanced DNA flexibility are consistent
with kinking as implied by the known thermodynamic and mechanical properties of
DNA. Further, as the model is generally applicable, oxDNA can be used to compare
strong bending in a variety of distinct systems. Here, in the first of a series
of papers investigating the biophysics of highly bent DNA, we consider a
``molecular vice'' \cite{fields_euler_2013}. In an accompanying paper we study
cyclization \cite{Harrison_cyclization}, and in an upcoming future publication
we consider minicircles \cite{Harrison_minicircles}.

In particular, in this paper, we first investigate the general thermodynamics
of the strongly bent regime of DNA. Using this insight, we corroborate the
results of recent ``molecular vice'' experiments \cite{fields_euler_2013},
which provide one of the most direct probes of strong DNA bending.
Specifically, we reproduce the experimental high-salt buckling transition and
approximate critical buckling length. We then directly probe the microscopic
configurations responsible for this transition, validating the experimental
interpretation of a kink-induced buckling transition.

\begin{figure*}
\includegraphics[width=16.5cm]{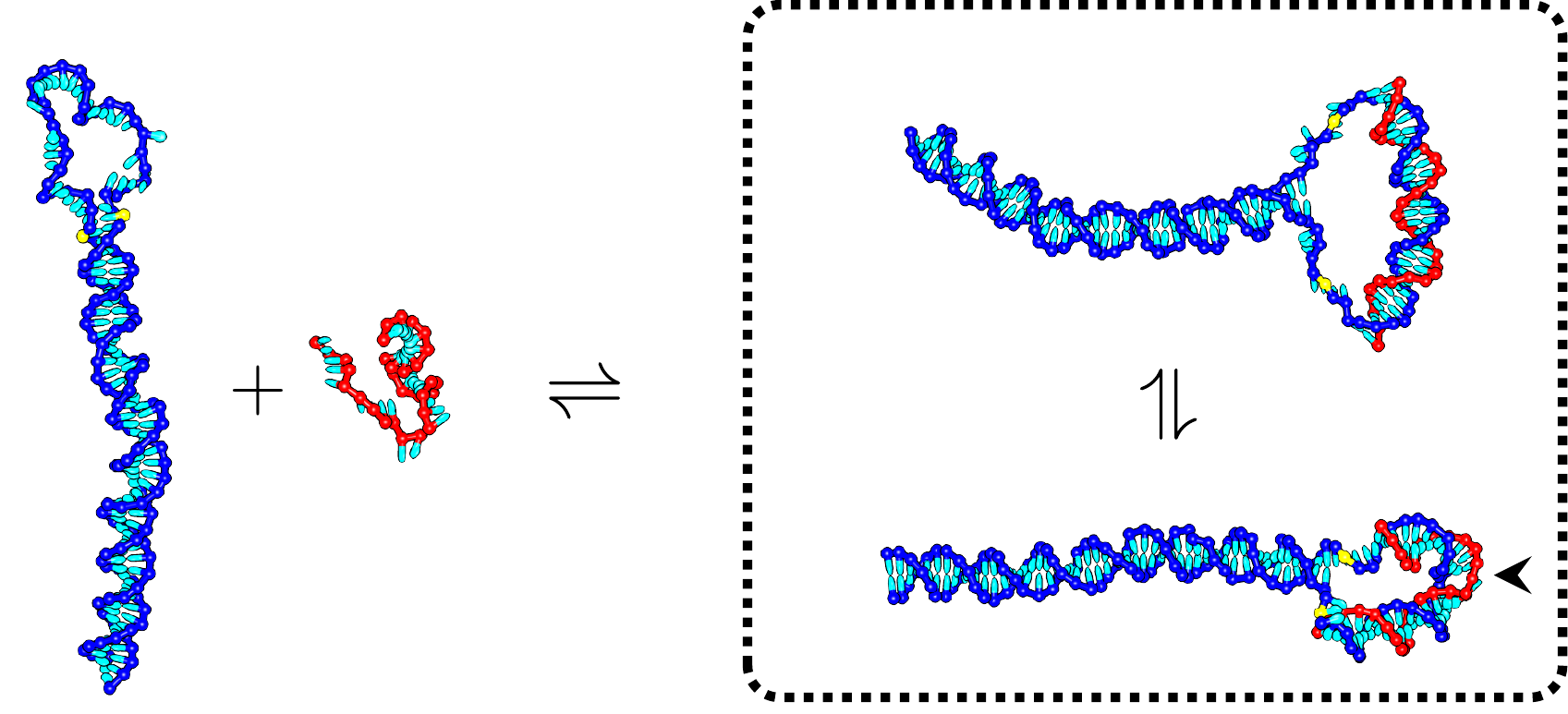}
\caption[]
{OxDNA representation of the ``molecular vice'' studied by Fields
\etalnoperiod\ \cite{fields_euler_2013}. A duplex of $N_\textnormal{dup}$ base
pairs is formed by hybridizing the $N_\textnormal{loop}$-nucleotide hairpin
loop with a variable-length complementary strand. The duplex may be kinked
({arrow}) or unkinked; in the latter case there is an equilibrium between continuous
bending in the duplex and unzipping in the stem. Fluorophores are attached to
stem-end thymines (\textcaption{yellow}) and are used for reporting hairpin
separation via FRET measurements.  For the representative system shown, the
hairpin stem is \SI{49}{} base pairs, $N_\textnormal{loop}={36}$ and
$N_\textnormal{dup}=\SI{30}{bp}$.  } 
\label{fig:cohen-diagram-system-oxDNA}
\end{figure*}

\section*{Materials \& Methods}
\subsection*{Concept of the ``molecular vice''}
In an elegant set of experiments, Fields \etal deployed a ``molecular vice'' to probe the strong bending
regime of DNA, providing evidence of a buckling transition {at high salt} that
cannot be explained by a WLC model \cite{fields_euler_2013}. They infer kinking
as the physical mechanism underlying this transition. The system they devised
consists of a DNA hairpin and complement strand
(\Cref{fig:cohen-diagram-system-oxDNA}); the hairpin has two regions, a loop
composed of $N_\textnormal{loop}$ nucleotides and a predominantly poly(AT) stem
composed of 49 base pairs (39 A--T base pairs adjacent to the loop and 10
terminal G--C base pairs). Complement strands of different lengths are
hybridized to the hairpin loop to form a duplex of $N_\textnormal{dup} <
N_\textnormal{loop}$ base pairs. The hairpin acts as a force clamp, imposing a
bending stress on the duplex. Formation of a duplex within the loop may require
part of the hairpin stem to be opened, which requires work against a salt-dependent
unzipping force. Stem unzipping is monitored via FRET (Fluorescence Resonance
Energy Transfer), reporting the separation between the fluorophores attached to
the first and last bases of the hairpin loop
(\Cref{fig:cohen-diagram-system-oxDNA}).

This setup is powerful because it allows the observation of stem unzipping as a
function of both the length of the complement strand and loop size. For
example, at short $N_\textnormal{dup}$, there is minimal stress on the system
-- it is necessary neither for the duplex to bend nor the stem of the hairpin
to unzip. At long $N_\textnormal{dup}$ (comparable to $N_\textnormal{loop}$),
the system cannot maintain full base-pairing without bending the duplex. {At
high salt Fields \etal observe for short loop lengths unzipping of the stem as
$N_\textnormal{dup}$ approaches $N_\textnormal{loop}$ base pairs. For some
longer loop lengths, they then see a sharp re-zipping when $N_\textnormal{dup}$
gets even closer to $N_\textnormal{loop}$ base pairs. Fields \etal argue that
the high-salt behaviour cannot be explained by the WLC, and infer the formation
of a sharp kink in the duplex that allows re-zipping \cite{fields_euler_2013}.}

\subsection*{The oxDNA model}
OxDNA \cite{ouldridge_structural_2011,ouldridge_coarse-grained_2012,sulc_sequence-dependent_2012} was developed to {bridge} the gap between all-atom molecular dynamics simulations and continuum or analytic approaches. It has been employed successfully for many systems, reviewed in reference \cite{doye_coarse-graining_2013}. Most importantly for this work, oxDNA has previously been shown to reproduce a range of structural transitions in response to applied stress \cite{matek_dna_2012, romano_coarse-grained_2013, matek_plectoneme_2014}.

The philosophy of the model is to encode the well-understood and extensively characterized aspects of DNA physics, such as its basic thermodynamic, mechanical and structural properties, into a model with physically-motivated interactions. Briefly, each nucleotide (sugar, phosphate and base group) is modelled as a rigid body with three interaction sites. A complex set of anisotropic interactions between the nucleotides have been parametrized to reproduce structural, thermodynamic and mechanical properties of single- and double-stranded DNA. The model is parameterized to a monovalent salt concentration of \ce{[Na+]=\SI{500}{\milli\Molar}}, where the electrostatic interactions are strongly screened. The thermodynamic, structural and dynamic properties of the model are fully characterized in references \cite{ouldridge_structural_2011} and \cite{ ouldridge_coarse-grained_2012}. These include basic mechanical properties such as persistence length and torsional rigidity.

The model has two parameterizations, average-base \cite{ouldridge_structural_2011,ouldridge_coarse-grained_2012} and sequence-dependent \cite{sulc_sequence-dependent_2012}. In the average-base parameterization, the base pairing and stacking interactions of each base are identical; whereas in the sequence-dependent parameterization \cite{sulc_sequence-dependent_2012}, stacking and hydrogen-bonding interactions vary by base identity (e.g.\ A--T base pairs weaker than G--C base pairs). We use the average-base parameterization to elucidate the fundamental thermodynamics of DNA bending since this illustrates the fundamental physics more clearly. Sequence-dependent thermodynamics are used for a more direct comparison with the molecular vice experiments and to elucidate sequence effects on the free-energy profile for bending.

\subsection*{Bending simulations}
Simulations to probe the free-energy cost of bending were performed with the virtual-move Monte Carlo (VMMC) algorithm (the variant in the appendix of Ref.\ \cite{Whitelam2009}) in the NVT ensemble at \SI{298}{\kelvin}. As the entire DNA bending regime is of interest, including strongly bent states that {are high in free energy}, umbrella sampling \cite{torrie_nonphysical_1977} was employed to efficiently sample the free-energy landscape of DNA bending. The end-to-end distance $R_\textnormal{ee}^\prime$ between the base interaction sites of the first and last bases along a strand was used as an order parameter for the umbrella sampling procedure, whereby the model Hamiltonian is augmented with an empirical potential that is iteratively adjusted to yield a roughly uniform sampling of the relevant values of $R_\textnormal{ee}^\prime$. The statistical bias of this artificial potential is removed at the analysis stage. Further details of simulation procedure are given in Supplementary \Cref{sec:cohen-methods-bending-simulations}.

To simplify interpretation, base pairing at the ends of the duplex  were enforced through the umbrella potential to explicitly disallow ``fraying'' of duplexes. Fraying is a possible mechanism for relaxation of stress in bent systems \cite{fields_euler_2013,kim15}, but we do not consider it in this section as {we are interested in the bending of the main body of the duplex, not its ends. We allow fraying in the molecular vice system, but, as discussed by Fields \etal \cite{fields_euler_2013}, it is} explicitly set up to inhibit fraying of the $N_\textnormal{dup}$ base pairs. Results are essentially identical for any constraint in which \SIrange{1}{5}{} nucleotides at either end of the duplex are constrained to be base paired (Supplementary \Cref{sec:cohen-no-fray}); 
in this article we quote results in which we constrain only the final base pairs. Imposing such a constraint allows us to sensibly define and measure the end-to-end distance between the centre of the helical axis, $R_\textnormal{ee}$, as the first and last base pairs are guaranteed to be present. Error bars in reported data represent the standard error of the mean from 5 independent simulations.

\subsection*{Molecular vice simulations}
Simulations of the molecular vice were conducted with the sequence-dependent parameterization of oxDNA using the VMMC algorithm in the NVT ensemble at \SI{25}{\celsius}. Error bars represent the standard error of the mean from 6 independent simulations. These simulations were performed without any biasing through the use of umbrella potentials. Further details of simulation procedure are given in Supplementary \Cref{sec:cohen-methods-molecular-vice-simulations}.

\subsection*{Definitions}

Our study investigates the role of several structural defects, the detection and reporting of which rely on practical definitions. Namely, we require definitions of fraying, bubble formation and kinking. The identification of these defects requires, in turn, a definition of a formed base pair. Following previous work \cite{ouldridge_structural_2011}, we define a base pair as formed if the hydrogen-bonding interaction between the two complementary bases is stronger than \SI{4.142e-21}{\joule} ($\sim \SI{1}{\kT}$ at $T=\SI{298}{\kelvin}$). We define fraying as the loss of base pairs at the duplex ends, while bubbles are defined as the opening of base pairs in otherwise continuous stretches away from the duplex ends.

A practical definition of a kink is less straightforward. Qualitatively, a kink is an area of pronounced curvature localized to a small region. There is some diversity of opinion on the quantitative description and classification of kinks \cite{mitchell_atomistic_2011}; however, they have in common the breaking of base pairs and/or stacking. Other requirements may include a change in the relative orientation of nucleotides. Due to the ambiguity in the definition of a kink, we explore both structural and energetic definitions. 

Structurally, we define a kink using the relative orientation of consecutive nucleotides. In oxDNA, the orientation of each nucleotide is unequivocally determined by two orthogonal unit vectors, the base-backbone vector and the base-normal vector. The base-backbone vector connects the backbone and base interaction sites of each nucleotide. The model is designed such that, in a relaxed duplex, the base-normal vector at index $i$, $\mathbf{\hat{a}}_i$, and at the consecutive index $i+1$, $\mathbf{\hat{a}}_{i+1}$, are approximately parallel; that is, $\mathbf{\hat{a}}_i \cdot \mathbf{\hat{a}}_{i+1} \approx 1$. A kink is defined to be present if $\mathbf{\hat{a}}_i \cdot \mathbf{\hat{a}}_{i+1} < 0$, a condition implying that the orientation of consecutive nucleotides {differs by more than \SI{90}{\degree}.}

Energetically, we define a kink as a disruption in stacking between adjacent base pairs, where the stacking potential energy is less than one-tenth its typical value in a duplex (\SI{4.142e-21}{\joule}). For the oxDNA model, the structural and energetic criteria for a kink behave very similarly when kinks are reasonably probable due to strong bending of the duplex, as shown in \Cref{fig:umbrella-free-energy-bending-D30}\,(c). In this regime, the detection of a kink is reasonably unambiguous and straightforward as favourable kinks are necessarily strongly bent, and both criteria do a reasonable job of identifying them.  At larger end-to-end distances, the two criteria give distinct results, and caution should be used when interpreting the data; a fuller discussion is provided in Supplementary \Cref{sec:cohen-methods-detectors}.

\section*{Results \& Discussion}

\subsection*{DNA bending}

\begin{figure}
\includegraphics[width=8.4cm]{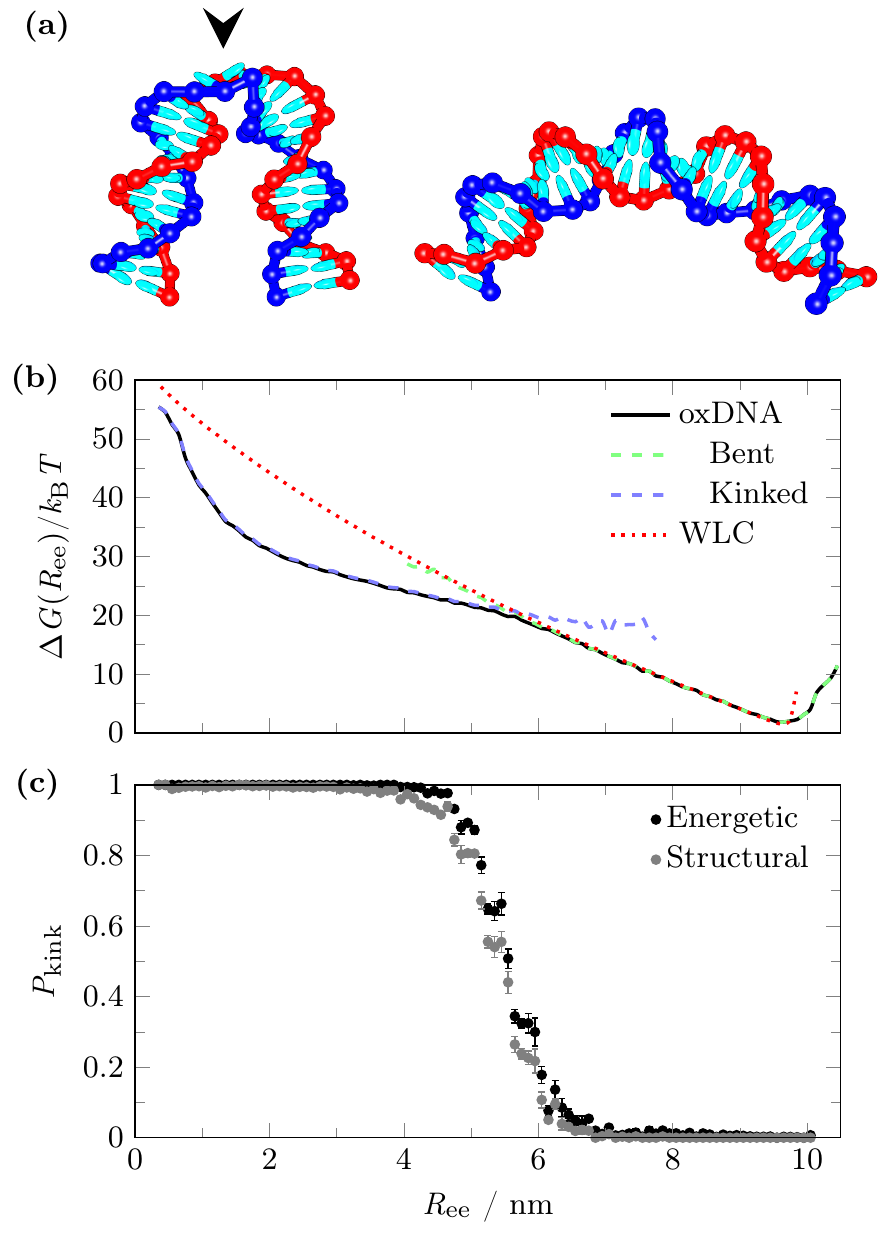}
\caption[]
{
\textbf{(a)} Representative OxDNA configurations for a {30}-{bp} duplex, depicting the kinked ({arrow}) and continuously bent states at $R_\textnormal{ee} = \SI{4}{\nano\meter}$ and $R_\textnormal{ee} = \SI{8}{\nano\meter}$, respectively.
\textbf{(b)} Free energy of bending $\Delta G(R_\textnormal{ee})$ for oxDNA ({\color{black}\textcaption{solid black}}), with the decomposed free energies of the continuously bent ({\color{black}\textcaption{light green dashed}}) and kinked ({\color{black}\textcaption{light blue dashed}}) states overlaid. OxDNA is compared to the WLC model approximation of Becker \etal \cite{becker_radial_2010} ({\color{black}\textcaption{red dotted}}). The comparison is fit-free, using known oxDNA parameters for persistence length (\SI{41.82}{\nano\meter}) and rise per base pair along the helical axis (\SI{0.34}{\nano\meter}) \cite{ouldridge_coarse-grained_2012} 
Note that, 
rather than setting the minimum $\Delta G(R_\textnormal{ee})$ to be zero as elsewhere in the text, we normalize $\Delta G(R_\textnormal{ee})$ so that $\exp{ \left( \sfrac{-\Delta G}{\SI{}{\kT}} \right)} = P(R_\textnormal{ee})$ in both cases.
\textbf{(c)} Probability of kinking $P_\textnormal{kink}$ using energetic (\textcaption{black})  and structural ({\color{black}\textcaption{grey}}) criteria for a kink, which are based upon disruption of stacking and base orientation, respectively. 
}
\label{fig:umbrella-free-energy-bending-D30}
\end{figure}

We start our discussion by characterising the basic thermodynamics of DNA bending in oxDNA. We simulate a {30}-{bp} DNA duplex and report the free-energy profile $\Delta G (R_\textnormal{ee})$ as a function of the end-to-end distance $R_\textnormal{ee}$  (\Cref{fig:umbrella-free-energy-bending-D30}\,(b)). The free energy of duplex bending has two distinct regimes, indicated by a change in slope at $R_\textnormal{ee}^\textnormal{trans} \approx \SI{5.5}{\nano\meter}$. The change in slope of the free-energy profile in \Cref{fig:umbrella-free-energy-bending-D30}\,(b) correlates with a crossover from continuously bent to kinked macrostates (\Cref{fig:umbrella-free-energy-bending-D30}(c)). Representative oxDNA structures for the continuously bent and kinked {30}-{bp} duplex (\Cref{fig:umbrella-free-energy-bending-D30}\,(a)) reveal that duplex kinking typically induces a \SIrange{1}{3}{bp} bubble at the kink ({further quantified in} Supplementary \Cref{sec:umbrella-nhist-bubble-D30})
and that this kink is located near the centre of the duplex (Supplementary \Cref{sec:D30-umbrella-nhist-kink-location}).

In \Cref{fig:umbrella-free-energy-bending-D30}\,(b) we also compare oxDNA
results to the WLC model;  specifically to the analytic  approximation to the
WLC end-to-end distance developed by Becker \etal \cite{becker_radial_2010}. 
This approach provides a good description of WLC behaviour across the
near-rigid, semiflexible and flexible regimes. As expected, agreement between
the WLC (when using the base pair rise and persistence length appropriate to
oxDNA) and oxDNA is excellent for intermediate $R_\textnormal{ee}$. 
The deviations most pertinent to this paper arise at $R_\textnormal{ee}\lesssim \SI{6}{nm}$ 
where the oxDNA model develops a kink that allows the system to lower its free energy 
compared to the WLC model. 
Deviations also arise 
at longer
$R_\textnormal{ee}$ near the DNA contour length $R_\textnormal{contour}$,
because the WLC neglects the finite intrinsic extensibility of DNA.
Additionally, the WLC does not account for the finite thickness of DNA, which
causes excluded volume effects at very short $R_\textnormal{ee}$, reported by
oxDNA as a sharp rise in $\Delta G$ at $R_\textnormal{ee} \approx
\SI{2}{\nano\meter}$.

\begin{figure}
\includegraphics[width=8.4cm]{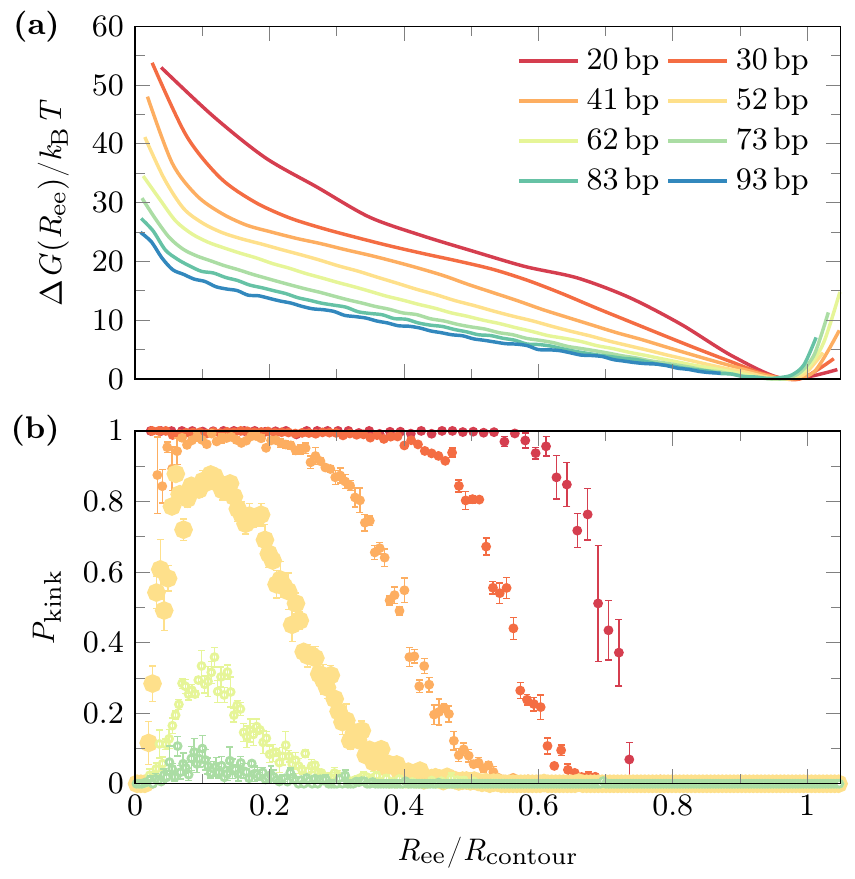}
\caption[]
{
\textbf{(a)} Length dependence of $\Delta G(R_\textnormal{ee})$, the duplex DNA bending free energy, as a function of reduced end-to-end distance $R_\textnormal{ee}/R_\textnormal{contour}$. Error bars are comparable to the line thickness.
\textbf{(b)} Length dependence of duplex DNA kinking. The structural criterion for kinking is shown; results using the energetic criterion are very similar in the kink transition region. 
}
\label{fig:umbrella-free-energy-bending-length}
\end{figure}

According to the WLC model, the enthalpic cost of bending a duplex to some
fraction of its contour length decreases with increasing duplex length. By
contrast, there is no obvious reason why the free-energetic cost of forming a
kink at the centre of the duplex should depend strongly on duplex length. We
therefore expect that, as duplex length is increased, the DNA will need to be
relatively more bent (smaller $R_\textnormal{ee}/R_\textnormal{contour}$) to
favour the formation of kinks. 
For sufficiently long DNA, we expect to see unkinked duplexes even for small
$R_\textnormal{ee}/R_\textnormal{contour}$. These expectations are confirmed in
\Cref{fig:umbrella-free-energy-bending-length}; long duplexes ($\ge
\SI{73}{bp}$) do not show any substantial kinking at any $R_\textnormal{ee}$, and for
shorter duplexes the length-dependence of the kink transition is as expected.
The re-entrant behavior where $P_\textnormal{kink}$ drops again at extremely small
$R_\textnormal{ee}/R_\textnormal{contour}$ arises from a change in geometry
that is enforced by excluded volume effects; for a detailed discussion see
Supplementary \Cref{sec:umbrella-diagram-small-ell-continious-bending}.

We next consider the question of what is the free-energy cost of kink
formation, a question which turns out not to be so easy to answer
unambiguously.  One natural approach is to define a free-energy cost $\Delta
G_\textnormal{trans}$ that corresponds to 
that at the middle of the kink transition, e.g.\ $\Delta G_\textnormal{trans} =
\Delta G (R_\textnormal{ee}^\textnormal{trans} ) = \SI{18}{\kT}$ for the
{30}-{bp} duplex in \Cref{fig:umbrella-free-energy-bending-D30}. This quantity
is essentially a measure of work that must be done on the duplex by pulling its
ends together before kink formation occurs. This numerical value is comparable
to the estimate of Fields \etal for ``$\Delta G_\textnormal{collapse}$'' 
\cite{fields_euler_2013}, the bending stress released upon kinking for a
similar system (in that kink formation is driven by pulling two ends of a
duplex together).  However, free energies must be carefully defined if
quantitative comparisons are to be made. In particular, neither $\Delta
G_\textnormal{trans}$ nor $\Delta G_\textnormal{collapse}$ is a generic
free-energy cost of kink formation that can be directly compared with other
systems in which kink transitions may occur. Furthermore, even when the bending
stress is applied in the same way, $\Delta G_\textnormal{trans}$ is not a
constant, but, as we show in Supplementary \Cref{tab:free-energy-kink-formation} for the
current system, is a function of duplex length (because the degree of bending
at the kink transition depends on length). 

One might instead define a quantity $\Delta G_\textnormal{kink}$ through the
probability of spontaneous kink formation (in the absence of stress) at a
specific location in the duplex. $\Delta G_\textnormal{kink}$ would in
principle be directly measurable from kinking probabilities of relaxed
duplexes. Although in principle this would be expected to be a generic
(sequence-dependent) quantity, its value will be strongly dependent on the
criteria used to identify kinks (Supplementary \Cref{sec:cohen-methods-detectors}).
This is in sharp contrast to
$\Delta G_\textnormal{trans}$, which is almost identical for both criteria used
to identify kinking in this work. In the presence of sharp bending, the
overwhelming majority of disruptions to the basic duplex structure are sharply
bent regions with broken stacking  interactions, which are unambiguously
``kinks''. In an unstressed system, however, the most common deviations from
the basic duplex are not unambiguously identifiable as kinks, making the
definition of a meaningful $\Delta G_\textnormal{kink}$ problematic.

Even given well-defined criteria for identifying kinks, $\Delta
G_\textnormal{kink}$ is not directly relatable to $\Delta
G_\textnormal{trans}$. Two reasons for this difference are particularly evident
in the context of bending duplexes. Firstly, to be a generic quantity, $\Delta
G_\textnormal{kink}$ must be defined through the probability that a kink occurs
that is centred on a specific base pair or (base pair step) in a duplex. In the
context of duplex bending, however, kink formation can occur over a finite
range of locations (
Supplementary \Cref{fig:umbrella-nhist-kink-location}), although kinks near to the centre
are preferred, and this fact is reflected in $\Delta G_\textnormal{trans}$.
Secondly, in pulling the ends of the duplex together to
$R_\textnormal{ee}^\textnormal{trans}$, we do work on both the unkinked {\it
and} the kinked state. Although kinks are more flexible than unkinked duplexes
they still have a finite resistance to bending, due both to entropic effects
and the need to strain the kink itself, as evidenced from the post-kinking
slope in $\Delta G(R_\textnormal{ee})$ in
\Cref{fig:umbrella-free-energy-bending-D30}(b). To compensate, we must apply
even more stress (than if the kink was completely flexible) to the system to reach the point at which kinks become
favourable relative to unkinked duplexes, tending to increase $\Delta
G_\textnormal{trans}$.

That the bending stress at which kinking occurs in a specific system ($\Delta G_\textnormal{trans}$) is not equivalent to the intrinsic cost of kink formation ($\Delta G_\textnormal{kink}$) emphasizes the difficulty in comparing between distinct experimental setups, in which the factors contributing to $\Delta G_\textnormal{trans}$ may well be different. For example, in some setups, it may be possible to relieve stress by kinking at one of a number of positions, as observed in simulations of short minicircles \cite{Harrison_minicircles}. In others, such as the molecular vice \cite{fields_euler_2013}, the kinking location may be constrained to be near the mid-point of a duplex. The free-energetic cost of bending the kinked state may also be distinct in different systems. 
Unless care is taken, such differences will hamper quantitative comparison of free energies. Analytic models \cite{du_kinking_2008, fields_euler_2013} used to extract free energies from data must therefore be carefully chosen if quantitative comparisons are to be drawn. Indeed, one of the advantages of explicit simulation with coarse-grained models is that direct comparisons can be made between distinct systems at the level of physical observables, rather than indirectly via free energies derived from analytic models. In this work, we will only make direct comparisons of $\Delta G_\textnormal{trans}$ for systems that are identical but for the DNA sequences, in which case differences will largely reflect differences in the intrinsic cost of kinking, $\Delta G_\textnormal{kink}$.

\begin{figure*}
\includegraphics[width=17.8cm]{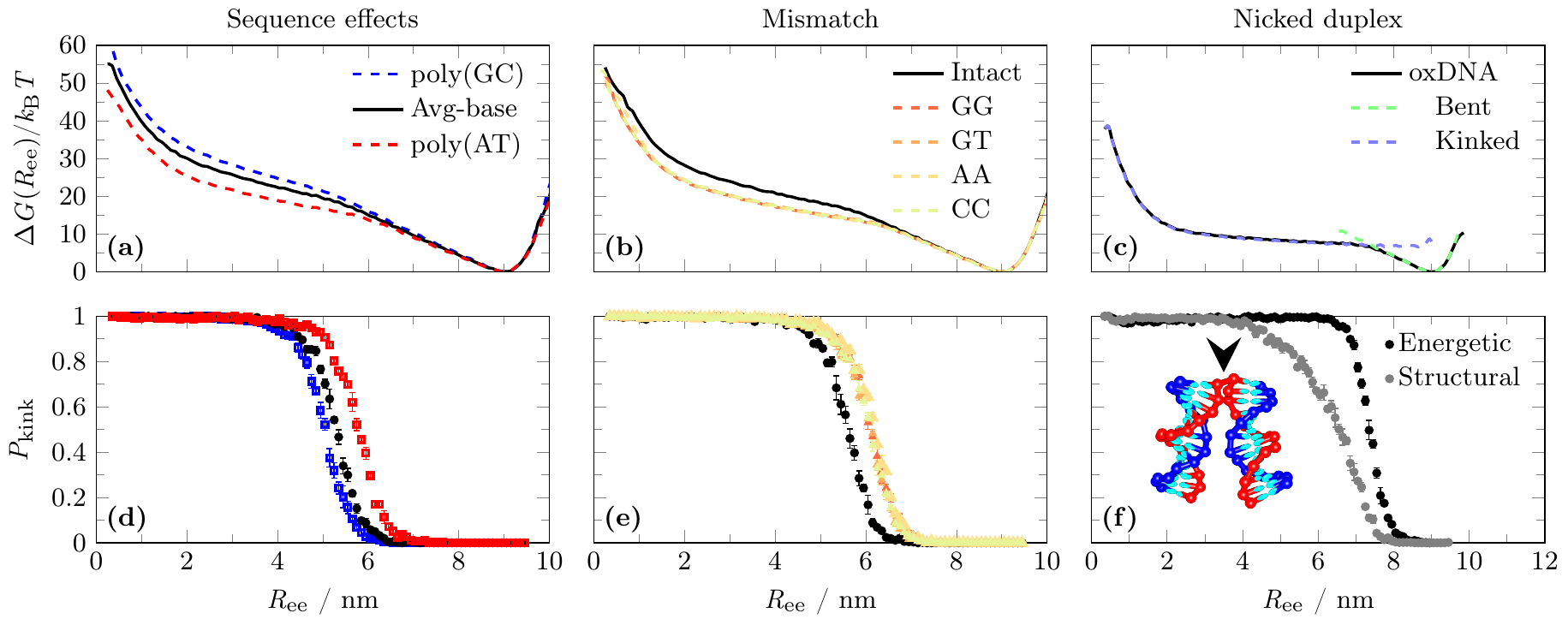}
\caption[]
{
\textbf{(a)} Sequence dependence of bending for a {28}-{bp} intact DNA duplex, comparing the sequence-dependent and average-base oxDNA parametrizations. Error bars are comparable to the line thickness.
\textbf{(b)} Impact of mismatches on the bending of a {28}-{bp} DNA duplex, compared to intact DNA, using the oxDNA sequence-dependent parameterization. Mismatches, ordered by increasing free-energy cost, are as follows: GG, GT, AA and CC. Sequences from Fields \etal \cite{fields_euler_2013}.
\textbf{(c)} Impact of a nick on the bending of a {28}-{bp} duplex for oxDNA ({\color{black}\textcaption{solid black}}), with the decomposed free energies of the continuously bent ({\color{black}\textcaption{light green dashed}}) and kinked ({\color{black}\textcaption{light blue dashed}}) states overlaid.  
\textbf{(d)} Sequence-dependence of kinking using the structural criterion for a kink; results using the energetic criterion are very similar.
\textbf{(e)} Impact of mismatches on kinking.
\textbf{(f)} Impact of a nick on kinking, via energetic (\textcaption{black}) and structural (\textcaption{grey}) criterions. A representative oxDNA configuration of a nicked duplex with a kink ({arrow}) at $R_\textnormal{ee}=\SI{4}{\nano\meter}$ is also depicted.
}
\label{fig:umbrella-free-energy-bending-sequence}
\end{figure*}

Given that the relative free energies of bending and kinking can also be perturbed with sequence variation \cite{geggier_sequence_2010}, we expect that sequence variation should affect the kink transition. We report sequence dependence in the free energy $\Delta G(R_\textnormal{ee})$ of bending for a {28}-{bp} duplex (\Cref{fig:umbrella-free-energy-bending-sequence}\,(a,d)). It is clear from these data that the poly(AT) sequence undergoes kinking more easily than the poly(GC) sequence; $\Delta G_\textnormal{trans}$  is lower, presumably due to the weaker base-pairing in the poly(AT) sequence. A secondary observation is that the poly(AT) sequence is slightly more flexible prior to the kink transition in oxDNA. These two effects should, in principle, have opposing consequences for $R_\textnormal{ee}^\textnormal{trans}$, tending to increase and decrease $R_\textnormal{ee}^\textnormal{trans}$ respectively. It is clear from \Cref{fig:umbrella-free-energy-bending-sequence}\,(d), however, that the ease of kinking dominates and poly(AT) has a longer $R_\textnormal{ee}^\textnormal{trans}$.

We also observe that a \SI{65}{\percent} GC sequence has a slightly lower $\Delta G_\textnormal{trans}$ than the average-base model (Supplementary \Cref{tab:free-energy-kink-formation}). This behavior is expected since the duplex will kink and bend preferentially in AT-rich regions. Due to the lack of easily kinkable regions, the average-base model is harder to kink than would be na\"{i}vely expected from an ensemble of \SI{50}{\percent} GC-content sequences.

Having investigated sequence variation, we now turn our attention to a system in which we introduce a mismatch, a structural motif that substitutes a base pair with a non-Watson-Crick pair. It is reasonable to expect that kinking will localize to the mismatch, provided it is conveniently positioned, and $\Delta G_\textnormal{trans}$ will be reduced. Following Fields \etal \cite{fields_euler_2013}, a mismatch was introduced at the mid-point of an otherwise intact duplex. We investigated GG, GT, AA and CC mismatches. 

As expected, oxDNA reports that mismatched sequences have a lower $\Delta G_\textnormal{trans}$ and undergo the kink transition at longer $R_\textnormal{ee}^\textnormal{trans}$ (\Cref{fig:umbrella-free-energy-bending-sequence}\,(b,e)). The change in $\Delta G_\textnormal{trans}$ is consistent with typical free energy costs associated with mismatches \cite{santalucia_thermodynamics_2004}: oxDNA reports \SI{16.1}{\kT} for the perfectly-matched duplex versus \SIrange{12.8}{13.5}{\kT} for the mismatches, a deviation of $\sim \SI{3}{\kT}$  (see \Cref{tab:free-energy-kink-formation}). Fields \etal  also found evidence that ease of kinking was correlated with the sequence-dependent mismatch cost predicted by the SantaLucia model of DNA thermodynamics \cite{fields_euler_2013}. 
OxDNA's treatment of mismatches is extremely simple, neglecting any residual interactions or possible steric clashes between geometrically incompatible bases. Consequently, it does not strongly distinguish between different mismatches, and does not reproduce sequence-dependent mismatch variation.

Like a mismatch, a nick, which is a break in an otherwise contiguous stretch of DNA whereby the backbone phosphodiester bond is severed, is a structural motif expected to impact DNA bending. In the relaxed state, base pairing is preserved and coaxial stacking may occur between the adjacent, but now non-covalently bonded, nucleotides. It is relatively easy for a nicked system to give sharp kinks whilst only disrupting stacking and not hydrogen-bonding (\Cref{fig:umbrella-free-energy-bending-sequence}\,(f)). We observe, unsurprisingly, that the presence of a nick dramatically reduces the bending free energy at the kink transition (\Cref{fig:umbrella-free-energy-bending-sequence}\,(c)). In accordance with this lower bending free energy ($\Delta G_\textnormal{trans} = \SI{6.7}{\kT}$), the crossover from continuously bent to kinked regimes
is shifted to larger $R_\textnormal{ee}^\textnormal{trans}$ by about $\SI{2}{\nano\meter}$.

It is also worth noting that DNA with a nick is far easier to bend then normal duplex DNA after the kink transition: the post-kinking slope for the nicked system (\Cref{fig:umbrella-free-energy-bending-sequence}\,(c)) is much shallower than for the intact duplex (\Cref{fig:umbrella-free-energy-bending-sequence}\,(b)). The absence of the constraint of the additional backbone bond in the nicked system makes extremely sharp bending very easy once coaxial stacking has been broken; this contrast highlights the work that must be done to strongly bend a kinked duplex without a nick. We note that for the nicked system, a more substantial difference between energetic and structural criteria for kinks can be seen (\Cref{fig:umbrella-free-energy-bending-sequence}\,(f)). This distinction arises because nicked systems undergo significant kinking at relatively large $R_\textnormal{ee}$, when it is less advantageous for kinked sections to bend extremely sharply; hence, because the structural criterion implies an effective angle $> \SI{90}{\degree}$ between consecutive bases, it can underestimate the prevalence of kinks.

\subsection*{Molecular Vice}
Having rigorously explored bending {in a generic context}, we turn to the ``molecular vice'' of Fields \etal \cite{fields_euler_2013}, which provides an exquisite experimental probe of the behaviour of duplex DNA under bending stress. We use oxDNA to explore the microscopic configurations that underlie their experimental observations. We will show that our findings are compatible with the observations of Fields \etal at high salt concentrations. In our simulations, the thymines located at the first and last bases of the loop were chosen as the reporter for hairpin separation since they are the closest analogues to the experimental setup. Specifically, we define $R_\textnormal{ee}^\textnormal{hairpin}$ as 
the separation between these bases' centres of mass.

\begin{figure}
\includegraphics[width=8.4cm]{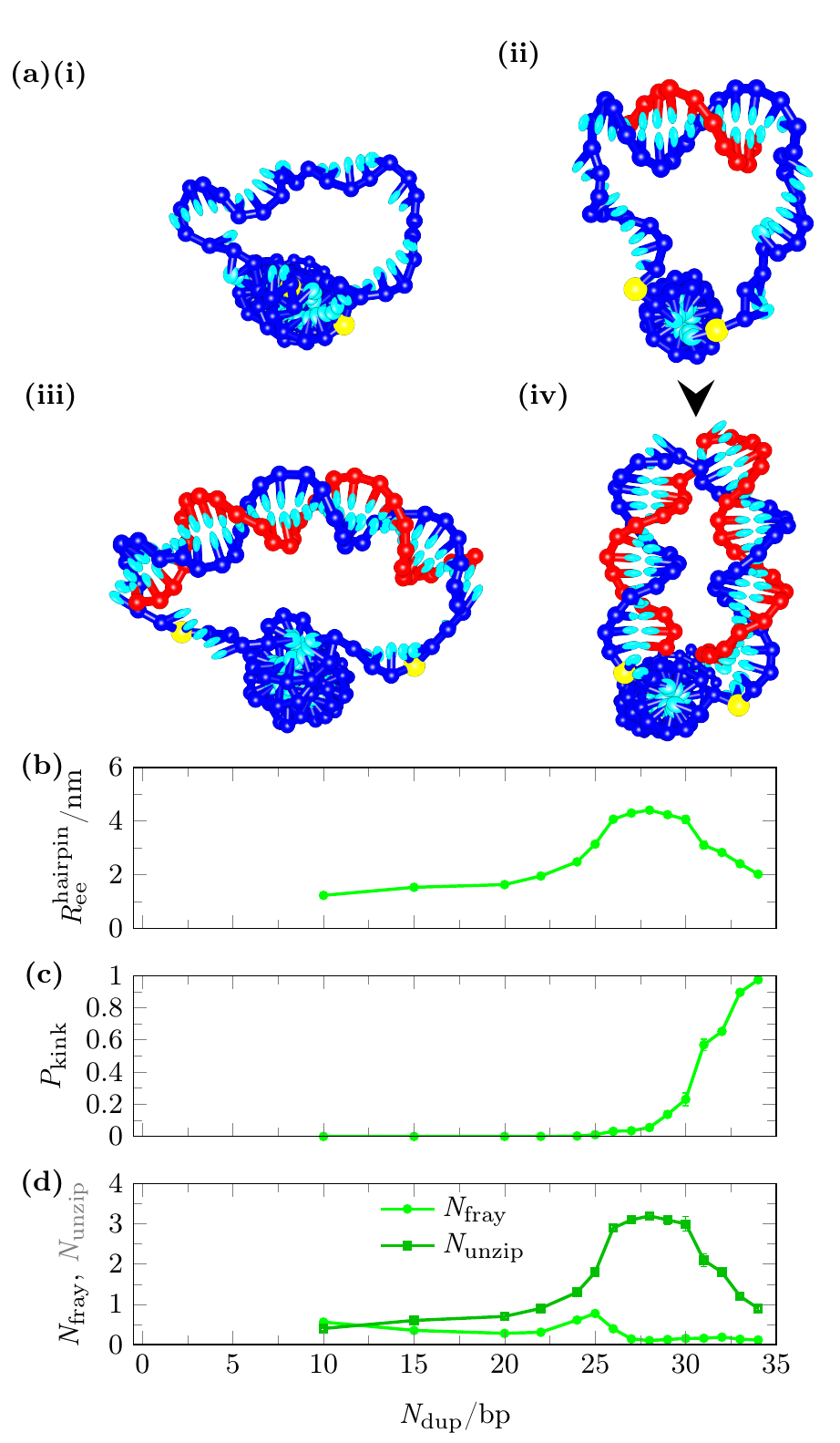}
\caption[]
{Structural properties of the molecular vice as modelled by oxDNA for $N_\textnormal{loop}=36$ and variable $N_\textnormal{dup}$. 
\textbf{(a)(i-iv)} Representative structures for $N_\textnormal{dup}=0,10,28,34$\,bp, respectively. 
Stem-end thymines (\textbf{yellow}) are used for fluorophore attachment and reporting hairpin separation. 
A kink is highlighted with an \textbf{arrow}.
\textbf{(b)} $R_\textnormal{ee}^\textnormal{hairpin}$ is the separation of the two thymines bases at the interface between the hairpin stem and loop. 
\textbf{(c)} $P_\textnormal{kink}$ is the probability of kinking in the duplex via the structural criterion.
\textbf{(d)} $N_\textnormal{fray}$ and $N_\textnormal{unzip}$ are the number of disrupted base pairs via fraying in the duplex (\textcaption{green circles}) and unzipping of the hairpin stem (\textcaption{dark green squares}), respectively. 
}
\label{fig:cohen-L36}
\end{figure}

We begin our discussion by considering the system with a loop length of $N_\textnormal{loop}=36$ nucleotides and a variable length duplex of $N_\textnormal{dup}$ base pairs (\Cref{fig:cohen-L36}). OxDNA reveals distinct behaviour for small, medium and large values of $N_\textnormal{dup}$. These behaviours are related to the most prevalent deformation modes from the canonical relaxed duplex DNA structure at each given $N_\textnormal{dup}$ length. Possible deformations to consider involve unzipping of the hairpin stem, and fraying, bending or kinking of the duplex.

\begin{figure}
\includegraphics[width=8.4cm]{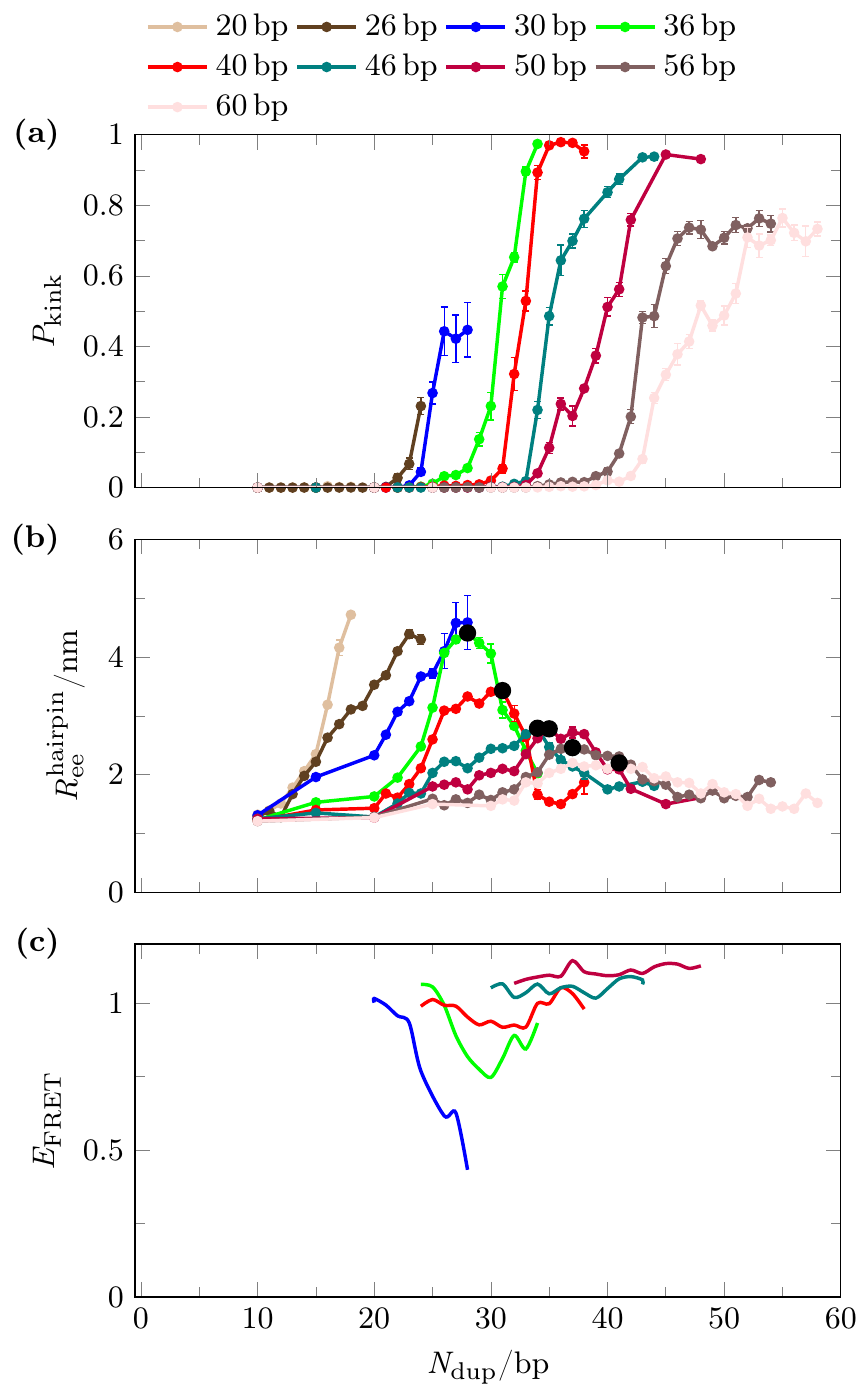}
\caption[]
{
Structural properties of the molecular vice for variable $N_\textnormal{loop}$.
\textbf{(a)} Kinking probability $P_\textnormal{kink}$ via the structural criterion.
\textbf{(b)} Hairpin separation $R_\textnormal{ee}^\textnormal{hairpin}$. Maxima in $R_\textnormal{ee}^\textnormal{hairpin}$ 
are highlighted (black circles).
\textbf{(c)} The experimental FRET efficiency $E_\textnormal{FRET}$ as reported in Ref.\ \cite{fields_euler_2013}. 
The available data is for $N_\textnormal{loop}=30,36,40,46,50$. 
}
\label{fig:cohen-Lall}
\end{figure}

For short duplexes ($N_\textnormal{dup} < \SI{20}{\bp}$) relative to the loop length $N_\textnormal{loop}=36$, the hairpin separation $R_\textnormal{ee}^\textnormal{hairpin}$  is comparable to that of a hairpin without an annealed complement strand (\Cref{fig:cohen-L36}\,(b)). \Cref{fig:cohen-L36}\,(c) and (d) show that kinking and fraying of the duplex, and hairpin unzipping, are small, and simulation snapshots such as \Cref{fig:cohen-L36}\,(a)(ii) do not show significantly bent duplexes. In this regime, the duplexes are too short to exert significant stress on the system.  

For intermediate-length duplexes ($N_\textnormal{dup}=\SIrange{20}{28}{\bp}$), $R_\textnormal{ee}^\textnormal{hairpin}$ increases due to stem unzipping (\Cref{fig:cohen-L36}\,(b)). The duplex is sufficiently long that the system cannot  maintain full base-pairing without experiencing considerable mechanical stress. The system must either bend  the duplex, or suffer a reduction in base pairing (either via unzipping the stem or fraying at the ends of the duplex). In fact, the system adopts configurations in which the deformation is shared between continuous bending of the duplex and unzipping, as can be seen from \Cref{fig:cohen-L36}\,(a)(iii). The bending energy released for each base pair unzipped decreases as the duplex gets straighter, whereas the cost of unzipping each base pair in the stem is approximately constant. The system can therefore minimize its free energy by adopting a state which combines unzipping with moderate bending. Unzipping the hairpin stem is favoured over fraying the duplex for two reasons; firstly, stem unzipping contributes two nucleotides worth of single-stranded DNA to the hairpin loop (thereby reducing the tension), versus a single base for fraying the duplex. Secondly, the weak poly(AT) stem is easier to melt than the strongly binding GC-rich regions at both ends of the duplex (this was a deliberate design choice of Fields \etal \cite{fields_euler_2013}).

For long duplexes ($28 < N_\textnormal{dup} \sim N_\textnormal{loop}$), $R_\textnormal{ee}^\textnormal{hairpin}$ decreases by re-zipping the stem and kinking, as reported in  \Cref{fig:cohen-L36}\,(b-d); a typical kinked structure is shown in \Cref{fig:cohen-L36}\,(a)(iv). Increasing $N_\textnormal{dup}$ increases the stress in the system, leading to stronger unzipping and bending costs if the duplex does not kink. The cost of kinking, however, is much less strongly dependent on $N_\textnormal{dup}$ (as was discussed in the general analysis of bending). Thus at sufficiently large  $N_\textnormal{dup}$, it is favourable to form a kink in the centre of the duplex, relieving the bending stress elsewhere and reducing the need to unzip. 

As well as considering the behaviour at fixed $N_\textnormal{loop} $, it is instructive to analyse the system as $N_\textnormal{loop}$ is varied, focusing on the highly stressed regime of $N_\textnormal{dup} \sim N_\textnormal{loop}$ (\Cref{fig:cohen-Lall}\,(a) and (b)). We observe three regimes: For small $N_\textnormal{loop} < 30$, unzipping and continuous bending are primarily observed, with almost no signal of kinking in $R_\textnormal{ee}^\textnormal{hairpin}$. For  $30<N_\textnormal{loop}< 60$, we see a sharp kink transition in which kinked states become dominant as $N_\textnormal{dup} \rightarrow N_\textnormal{loop}$ {($N_\textnormal{loop}=30$ sits near the crossover between these regimes)}. For larger $N_\textnormal{loop}$, this transition is less pronounced, eventually fading away. Additional simulations at $N_\textnormal{loop} = \SIrange{66}{90}{}$, with $N_\textnormal{dup} = N_\textnormal{loop}-2$, show the kinking probability decaying from $\sim \SI{30}{\percent}$ to $\sim \SI{2}{\percent}$ (Supplementary 
\Cref{fig:cohen-Lall-long}).

Why does a kink transition occur only for a finite range of $N_\textnormal{loop}$? Unzipping of base pairs is extremely effective at relaxing stress for short duplexes: each unzipped base pair opened allows the duplex to straighten substantially. Thus at short lengths, stress is primarily relaxed by unzipping, duplex bending is small and kinking does not come into play. As the duplex gets longer, each unzipped base pair has less of a relaxing effect on the duplex, as it provides a smaller contribution relative to the overall length discrepancy between the intrinsic length of the duplex and the end-to-end distance enforced by the vice. Consequently, the bending of the duplex is enhanced, and eventually a kink transition is observed when this bending is sufficiently severe. For longer duplexes, the enthalpic cost of continuously bending to satisfy the separation imposed by the vice drops with duplex length (this can be shown directly using the WLC expression of Becker \etal \cite{becker_radial_2010}), whereas kink formation does not get easier. Thus the driving force for kinking is reduced, and  kinking is absent at sufficiently long loop lengths.

Another notable point is that the maximum in $\langle R_\textnormal{ee}^\textnormal{hairpin}\rangle$ prior to the kink transition (indicated by black dots in \Cref{fig:cohen-L36}\,(b)) drops with increasing $N_\textnormal{loop}$ at values of $N_\textnormal{loop}$ for which kinking is observable. Why should this be? As mentioned above, unzipping of stem base pairs is increasingly ineffective for larger systems, and bending of the duplex becomes increasingly easy. Therefore the degree to which the system unzips as $N_\textnormal{dup} \rightarrow N_\textnormal{loop}$ is reduced for larger systems, and $\langle R_\textnormal{ee}^\textnormal{hairpin}\rangle$ does not grow as significantly prior to kinking.

We now compare the oxDNA results to those of Fields \etal  \cite{fields_euler_2013}. Unlike oxDNA, Fields \etal could not directly observe kinked macrostates. Instead, they inferred kinking from the non-monotonic behavior of the FRET signal between dyes attached to the first two bases of the loop (a proxy for unzipping - \Cref{fig:cohen-L36}\,(c)). They defined a critical buckling length at high-salt, which they interpreted as the onset of localized bending via kinking, as a local minimum in the FRET signal. We can define an analogous quantity in oxDNA, the maximum in $\langle R_\textnormal{ee}^\textnormal{hairpin} \rangle$ for a given $N_\textnormal{loop}$. 

In their higher salt experiments, Fields \etal saw evidence of unzipping and no kinking for $N_\textnormal{loop}={30}$, identified non-monotonic behavior of FRET for $N_\textnormal{loop}=36,40$ and saw no substantial change in the FRET signal with $N_\textnormal{dup}$ for larger values of $N_\textnormal{loop}$. Broadly speaking, these three regimes are consistent with our observations: kink transitions occur for intermediate values of $N_\textnormal{loop}$, with pure unzipping at smaller values and no unzipping or kinking at larger values. Quantitatively, we observe the beginning of a kink transition for $N_\textnormal{loop} = 30$, slightly earlier than Fields \etalnospace, and see evidence of kinking at larger values of $N_\textnormal{loop}$. The $N_\textnormal{dup}$ values at which kink transitions occur for $N_\textnormal{loop}={36}$ and $N_\textnormal{loop}={40}$ in oxDNA are within two base pairs of those inferred by Fields \etalnospace, as can be seen in \Cref{fig:cohen-L36}\,(c) (clearly, however, inferring the `exact' onset of kinking is not trivial from the experimental data). 

The quantitative consistency of results may be even stronger than immediately apparent. OxDNA is parametrized at \ce{[Na+]=\SI{500}{\milli\Molar}}, meaning that the stem of the vice should be somewhat more resistant to unzipping than in the experiments, which were performed at \ce{[Na+]=\SI{250}{\milli\Molar}}, perhaps explaining the onset of kinking for $N_\textnormal{loop}=30$ that is not apparent in the work of Fields \etal (for additional discussion, please see Supplementary \Cref{sec:cohen-temperature-L30})
Consistent with this argument, Fields \etal do not find evidence for kinking at all at much lower salt concentrations, when the stem of the vice is less stable. Slightly smaller values of $N_\textnormal{dup}$ at which buckling occurs for a given $N_\textnormal{loop}$ in oxDNA may also be explicable in this way.

Our results for larger values of $N_\textnormal{loop}$ also suggest that although kinking occurs, it does so following an increasingly small amount of unzipping. This is actually consistent with the observations of Fields \etal --- compare the drop in FRET prior to kinking for $N_\textnormal{loop}=36$ and $N_\textnormal{loop}=40$. We posit that unzipping prior to kinking may be undetectable through the methodology of  Fields \etal for $N_\textnormal{loop} \gtrsim 40$.
There is a highly non-linear relationship between FRET signal and fluorophore
separation, which is roughly equivalent to
$R_\textnormal{ee}^\textnormal{hairpin}$. FRET is most sensitive near the
F\"{o}rster radius, $R_\textnormal{0}=\SIrange{5.7}{6.7}{\nano\meter}$ for the
Cy3B/Alexa647 fluorophore pair used to probe stem unzipping. OxDNA reports
$\langle R_\textnormal{ee}^\textnormal{hairpin} \rangle < \SI{3}{\nano\meter}$
for $N_\textnormal{loop} \geq 46$, suggesting that it is possible that kinking
occurs in experiment at larger values of $N_\textnormal{loop}$ without
sufficient change in FRET signal for it to be observable. OxDNA does suggest,
however, that kinking is never observed when  $N_\textnormal{loop}$ is large enough
that $N_\textnormal{dup} \sim N_\textnormal{loop}$ duplexes undergo continuous
bending. 

\section*{Conclusion}

In this paper we have provided a full characterization of the thermodynamics of
DNA bending in a coarse-grained model, oxDNA. When weakly bent, oxDNA is
well-described by the worm-like chain (WLC) model. However, if a sufficiently
strong degree of bending is imposed, oxDNA tends to adopt an alternative
conformation in which the strain is localized in a small region of disrupted
base-pairing and stacking interactions. The cost of such local kinks are
compensated by the relaxation of the rest of the molecule.  Such alternative
modes of relaxation are not captured by the WLC model and lead to an enhanced
probability of sharp bends. Although a kink is far more flexible than canonical
B-DNA, bending is not completely free. Conformations are more restricted than
for a duplex with a nick, for example, which is not constrained by two
continuous backbones.  

Our results enable us to identify the end-to-end constraint that is sufficient
to cause kinking (or localized bending). Short duplexes exhibit localized
bending at relatively larger end-to-end distance than long duplexes, while
sufficiently long duplexes ($\gtrsim \SI{70}{bp}$) do not undergo kinking even
if the ends are held as close together as possible; they are long enough to
achieve such a conformation through reasonably gentle bending distributed
throughout the molecule.  Additionally, AT-rich sequences exhibit localized
bending at longer end-to-end distance than GC-rich sequences, due to the lower
cost of disrupting weaker interactions. 

Our analysis of kinking reveals that one must be extremely careful in inferring free energies of kinking from specific systems. A natural definition of $\Delta G_\textnormal{kink}$ would be through the probability of spontaneous kink formation in relaxed DNA at a specific site -- this should be a generic property depending on the local sequence, and would be the quantity measurable if relaxed duplexes were examined for kinking. 
If kinking is enforced through external bending of duplex DNA, however, the amount of work that must be done on the duplex state to enforce a transition, $\Delta G_\textnormal{trans}$, is potentially quite distinct from $\Delta G_\textnormal{kink}$. Reasons include the fact that work may also be done in bending the kinked state to reach the point of transition, and the fact that specific experimental systems (such as the molecular vice \cite{fields_euler_2013}) may favour kink formation at  certain locations whereas other systems (such as DNA minicircles \cite{Harrison_minicircles}) may allow kinks anywhere. These considerations make direct comparisons of free-energies of kinking as inferred through simple analytic models difficult, hindering the comparison of kinking in distinct experimental systems.

Computational models such as oxDNA have the virtue that they allow comparison
between different experimental systems in which kinking has been hypothesized,
as they can be directly simulated with the same model and experimentally
comparable observables can be computed. {In particular, physical factors such
as the work done in bending a kink, or the preference for some kink locations
over others in an experimental setup, should be naturally incorporated (at
least at a semi-quantitative level)}. Coarse-grained models parametrized to
reproduce basic DNA mechanics and thermodynamics therefore offer a valuable
tool for establishing whether proposed observations of kinking are consistent
with each other and our current knowledge of DNA.

To this end, we have  simulated the ``molecular vice'' of Fields \etal
\cite{fields_euler_2013}, a direct probe of DNA bending. {In this experiment,
the hybridization of a strand to a hairpin loop can lead to a range of
behaviours interpreted by  Fields \etal as duplex bending, kinking and hairpin
stem unzipping dependent on the length of the strand and hairpin loop.} Our
results are consistent with the observations of Fields \etalnospace, {support
their interpretation that kinking is responsible for rezipping at high salt},
and provide a more detailed mechanistic perspective on the data. We also
propose that duplex kinking might be present for longer duplexes than can be
directly inferred from the experimental data, due to the limits of FRET
precision and minimal unzipping prior to kinking.

OxDNA is a simplified model, and the level of agreement with Fields \etal
should not be over-interpreted. Although reproduction of the data is evidence
that the model is well-founded, a number of modelling simplifications, and the
slight difference in salt conditions in experiment and oxDNA, might give rise
to small quantitative deviations. 
The important point is that the general agreement between the experimental data
and oxDNA helps to further validate Fields \etal's inferral of kinking, and to
show that this behaviour is consistent with the known thermodynamics and
mechanics of DNA as captured by the oxDNA model.  Furthermore, we have recently
also performed careful comparisons between oxDNA simulations and experiments on
DNA cyclization \cite{Harrison_cyclization} and minicircles
\cite{Harrison_minicircles}.  
Reasonable consistency with experiment in these studies allows us to conclude
that the experimental evidence of kinking in these systems is broadly
consistent, and gives us further confidence that oxDNA describes the basic
physics of kinking well. This consistency is particularly important given the
controversy surrounding the existence of {kinks and enhanced DNA flexibility}
\cite{le_probing_2014, mazur_dna_2014}. We note that our observation that
kinking may go undetected for intermediate loop sizes in the setup of Fields
\etal is potentially important in this regard; otherwise the absence of kinking
for fairly short $N_\textnormal{loop}$ might be surprising when compared to
digestion assays on minicircles \cite{du_kinking_2008,demurtas_bending_2009},
particularly as simple mechanical arguments suggest that the geometry of
torsionally relaxed minicircles is less conducive to kinking than the geometry
of the molecular vice.

%\section*{Supplementary Data}
%Supplementary Data are available at NAR Online.

\section*{Funding}
This work was supported by the Engineering and Physical Sciences Research Council [EP/I001352/1], the National Science Foundation Graduate Research Fellowship Program, the National Institutes of Health National Heart, Lung and Blood Institute, Wolfson College, Oxford, and University College, Oxford.

\textit{Conflict of interest statement}. None declared.

\section*{Acknowledgements}
The authors acknowledge the use of the computing facilities of Oxford Advanced Research Computing and 
the e-Infrastructure South IRIDIS High Performance Computing Facility.

\clearpage

%\end{document}

%\input{paper-preamble}
%\externaldocument[M-]{HarrisonRM-NAR2014-Cohen} 

%\begin{document}

%\setcounter{equation}{0}
%\setcounter{figure}{0}
%\setcounter{table}{0}
%\setcounter{page}{1}
\makeatletter
\renewcommand{\bibnumfmt}[1]{[S#1]}
\renewcommand{\citenumfont}[1]{S#1}

\begin{widetext}
\begin{center}
\textbf{\large 
Supplementary material for ``Coarse-grained modelling of strong DNA bending I: Thermodynamics and comparison to an experimental molecular vice''
}
\end{center}
\end{widetext}

%\author{Ryan M. Harrison}
%\affiliation{Physical \& Theoretical Chemistry Laboratory, Department of Chemistry, University of Oxford, South Parks Road, Oxford, UK, OX1 3QZ}

%\author{Flavio Romano}
%\affiliation{Physical \& Theoretical Chemistry Laboratory, Department of Chemistry, University of Oxford, South Parks Road, Oxford, UK, OX1 3QZ}

%\author{Thomas E. Ouldridge}
%\affiliation{Department of Mathematics, Imperial College, 180 Queen's Gate, London, UK, SW7 2AZ}
%\affiliation{Rudolf Peierls Centre for Theoretical Physics, Department of Physics, University of Oxford, 1 Keble Road, Oxford, UK, OX1 3NP}

%\author{Ard A. Louis}
%\affiliation{Rudolf Peierls Centre for Theoretical Physics, Department of Physics, University of Oxford, 1 Keble Road, Oxford, UK, OX1 3NP}

%\author{Jonathan P. K. Doye}
%\affiliation{Physical \& Theoretical Chemistry Laboratory, Department of Chemistry, University of Oxford, South Parks Road, Oxford, UK, OX1 3QZ}

%\date{\today}

%\maketitle

\beginsupplement

%\FloatBarrier
%\tableofcontents

\section{Extended Methods}
\subsection{Bending simulations}
\label{sec:cohen-methods-bending-simulations}
Bending simulations were performed in three phases: exploratory, equilibration and production. In the exploratory phase, we iteratively adjusted the umbrella sampling biasing potential to yield a flat probability distribution $P(R_\textnormal{ee}^\prime)$, where $R_\textnormal{ee}^\prime$ is the end-to-end distance between the base interaction sites of the first and last bases on a single strand using a discrete potential with increments of \SI{0.4259}{\nano\meter}. As stated in the main text, the duplexes were constrained not to fray, allowing us to sample in terms of $R_\textnormal{ee}$, the end-to-end distance between the centres of the helical axis at the first and last base pairs, choosing an $R_\textnormal{ee}$ bin-width of \SI{0.085}{\nano\meter}. Non-native base-pairing was also forbidden. As the iteration was performed manually, simulation times varied in the range \SIrange{E6}{E7} VMMC steps per particle. 

In the equilibration phase, we equilibrated the system for \SI{E7}{} VMMC steps per particle using the aforementioned umbrella biasing potential. Adequate sampling (number of transitions between end-to-end distance bins) and decorrelation (via a block average decorrelation method) of potential energy, bubble size, fraying (where applicable) and structural kinking were checked. For the production phase, each production simulation (five independent simulations per measurement) was initialized with a statistically independent starting configuration and a unique random seed. This is accomplished by randomly drawing starting configurations from the equilibration phase, one for each production run, ignoring the first \SI{2E6}{} VMMC steps per particle of equilibration. The production run time for each simulation was \SI{E8}{} VMMC steps per particle. For reference, the characteristic decorrelation time for the potential energy is $\sim \SI{E4}{}$ VMMC steps per particle, while the decorrelation time for kinking is $\sim \SI{E5}{}$ VMMC steps per particle. 
The ``seed moves'' used to build clusters in the VMMC algorithm were:
\begin{itemize}
\item Rotation of a nucleotide about its backbone site, with an axis chosen uniformly on the unit sphere, and with an angle drawn from a normal distribution with a mean of zero and a standard deviation of 0.10 radians. 
\item Translation of a nucleotide, where the displacement along each Cartesian axis is drawn from a normal distribution with a mean of zero and a standard deviation of \SI{0.85}{\angstrom}. 
\end{itemize}

Additional simulations were performed to measure the free energy of kinked states for a {30}-{bp} duplex at larger values of $R_\textnormal{ee}$. In this case, as well as biasing $R_\textnormal{ee}$, the umbrella potential was used to forbid states without kinks (separate simulations were performed for each kink criterion). Umbrella potentials for $R_\textnormal{ee}$ were obtained via manual iteration at a preliminary stage. Simulations were started from kinked configurations and thermalized for \SI{3.3E6}{} VMMC steps per particle. Subsequent data collection lasted for \SI{3.3E8}{} VMMC per particle. Five independent simulations were performed. The VMMC ``seed moves'' differ slightly from those mentioned previously:
\begin{itemize}
\item Rotation of a nucleotide about its backbone site, with the axis chosen from a uniform random distribution and the angle from a normal distribution with mean of zero and a standard deviation of 0.15 radians;
\item Translation of a nucleotide with the direction chosen from a uniform random distribution and the distance from a normal distribution with mean of zero  and a standard deviation of \SI{1.28}{\angstrom}. 
\end{itemize}

\subsection{Constraints on fraying}
\label{sec:cohen-no-fray}
\begin{figure}[h!]
\includegraphics[width=8.5cm]{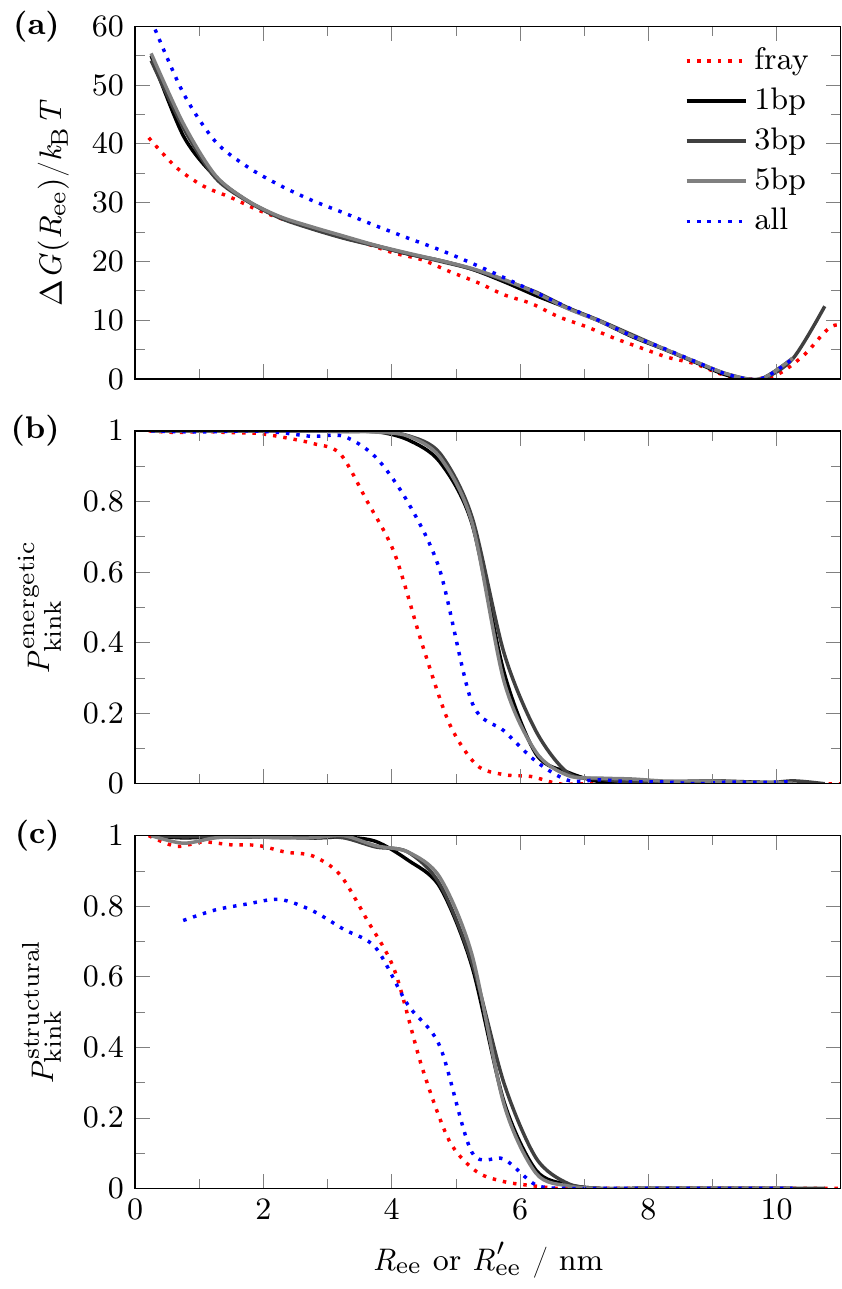}
\caption[]
{
The impact of constraining the terminal 1, 3 and 5 base pairs from fraying on the free energy of bending $\Delta G(R_\textnormal{ee})$ and kinking in an $N_\textnormal{bp}=\SI{30}{\bp}$ duplex. Limiting cases allowing fraying and enforcing hydrogen bonding along the entire duplex are also shown. \textbf{(a)} Free energy of bending. Probability of kinking based on the \textbf{(b)} energetic or \textbf{(c)} structural criteria for kinking. In all panels, for the frayed system $R_\textnormal{ee}$ (the end-to-end distance between the centres of the helical axis at the first and last base pairs) is ill-defined, so the very similar $R_\textnormal{ee}^\prime$ (the end-to-end distance between the base interaction sites of the first and last bases along the umbrella sampled strand) is used instead. 
}
\label{fig:cohen-no-fray}
\end{figure}

To isolate the effects of DNA bending, as opposed to alternative deformation modes such as fraying of the terminal base pairs, we impose a condition preventing fraying in our DNA bending simulations. {Note that this constraint was not applied in simulations of the molecular vice system.} Using an additional order parameter, the {30}-{bp} system was constrained such that the terminal base pairs remained hydrogen-bonded, thereby eliminating the possibility of fraying at duplex ends. We compared constraints where 1, 3 and 5 base pairs at the duplex ends were prevented from opening, finding that the various conditions yield comparable results (\Cref{fig:cohen-no-fray}). 

To further probe the behaviour of the {30}-{bp} system, in separate simulations we allowed the system to fray freely or preserved hydrogen bonding along the entire duplex. Unsurprisingly, {as frayed base pairs can adopt a large range of conformations including many that reduce $R_\textnormal{ee}$,  allowing fraying as an additional deformation mode reduces the bending free-energy penalty and delays the onset of kinking. Preserving hydrogen bonding throughout the duplex also delays kinking, but by limiting kinked states to those that do not involve disrupted base pairing; only those that break stacking but not base pairing remain, and thus the free energy cost of kinking is higher.}

Since a system in which fraying is permitted may not have terminal base pairs, the usual definition of $R_\textnormal{ee}$ reporting the separation between the helical axis centres of the terminal base pairs, is not well defined. Therefore, for this system we instead use $R_\textnormal{ee}^\prime$, defined as the separation between the base interaction sites of the first and last bases along a single strand.

\subsection{Molecular vice simulations}
\label{sec:cohen-methods-molecular-vice-simulations}
Akin to the bending simulations of duplex DNA (\Cref{sec:cohen-methods-bending-simulations}), the molecular vice simulations were performed in three phases: assembly, equilibration and production. An exploratory phase is not required since the simulations were unbiased, i.e.\ umbrella sampling is not used. As in the bending simulations, non-native base pairing was disallowed for simplicity. In the assembly phase, the system was initialized as two single strands. Harmonic traps were used to bring the stem ends, as well as the loop and complement strand, into proximity. Assembly with traps requires $\sim \SI{5E5}{}$ VMMC steps per particle, significantly faster than free assembly.

The assembled systems were equilibrated for \SI{E7} VMMC steps per particle. For the production phase, each production simulation (six independent simulations per point) was initialized with a statistically independent starting configuration and a unique random seed. This was accomplished by randomly drawing starting configurations from the equilibration phase, one for each production run, ignoring the first \SI{2E6}{} VMMC steps per particle of the equilibration phase. The length of each production run was $\sim \SI{E8}{}$ VMMC steps per particle.

All simulations were performed with the sequence-dependent oxDNA parameterization. Where available ($N_\textnormal{loop}=30,36,40,46,50$), the sequences are identical to those used in the experiments of Fields \etal \cite{Sfields_euler_2013}; where not available 
, similar sequences are used. All sequences are written $5^\prime$ to $3^\prime$.

The Fields \etal stem sequence is used for all systems:
\texttt{GCC CGG CGG CTT ATA AAA TTT ATT AAT TAT ATA TTT TAT TTA ATA TAA T-Loop}. A complete list of loop sequences is available in \Cref{tab:cohen-methods-sequences}.

\begin{table*}
\ra{1.3}
\begin{tabular}{ ll }
\toprule
\multicolumn{1}{c}{$N_\textnormal{loop}$} & \multicolumn{1}{c}{Loop sequence} \\
\midrule
20                          & \scriptsize{\texttt{{\color{red}T}AC CGA TAA GCT TGG TCA T{\color{red}T}}} \\
26                          & \scriptsize{\texttt{{\color{red}T}CC CAC CGA TAA GCT TGG TCA TGC C{\color{red}T}}} \\
30\superscript{$\ddagger$}\superscript{*} & \scriptsize{\texttt{{\color{red}T}CG CCC ACC GAT A{\color{blue}AG CT}T GGT CAT GCC CG{\color{red}T}}} \\
36\superscript{$\ddagger$}  & \scriptsize{\texttt{{\color{red}T}GC CCG CCC ACC GAT AAG CTT GGT CAT GCC CGC CG{\color{red}T}}} \\
40\superscript{$\ddagger$}  & \scriptsize{\texttt{{\color{red}T}CC GCC CGC CCA CCG ATA AGC TTG GTC ATG CCC GCC GCC{\color{red}T}}} \\
46\superscript{$\ddagger$}  & \scriptsize{\texttt{{\color{red}T}CC GCC GCC CGC CCA CCG ATA AGC TTG GTC ATG CCC GCC GCC GCC {\color{red}T}}} \\
50\superscript{$\ddagger$}  & \scriptsize{\texttt{{\color{red}T}GC CCG CCG CCC GCC CAC CGA TAA GCT TGG TCA TGC CCG CCG CCG CCC G{\color{red}T}}} \\
56                          & \scriptsize{\texttt{{\color{red}T}GC CGC CCG CCG CCC GCC CAC CGA TAA GCT TGG TCA TGC CCG CCG CCG CCC GCC G{\color{red}T}}} \\
60                          & \scriptsize{\texttt{{\color{red}T}CC GCC GCC CGC CGC CCG CCC ACC GAT AAG CTT GGT CAT GCC CGC CGC CGC CCG CCG CC{\color{red}T}}} \\
66                          & \scriptsize{\texttt{{\color{red}T}CG CCC GCC GCC CGC CGC CCG CCC ACC GAT AAG CTT GGT CAT GCC CGC CGC CGC CCG CCG CCC GC{\color{red}T}}} \\
70                          & \scriptsize{\texttt{{\color{red}T}GC CGC CCG CCG CCC GCC GCC CGC CCA CCG ATA AGC TTG GTC ATG CCC GCC GCC GCC CGC CGC CCG CCG {\color{red}T}}} \\
76                          & \scriptsize{\texttt{{\color{red}T}CC GGC CGC CCG CCG CCC GCC GCC CGC CCA CCG ATA AGC TTG GTC ATG CCC GCC GCC GCC CGC CGC CCG CCG GCC {\color{red}T}}} \\
80                          & \scriptsize{\texttt{{\color{red}T}GC CCG GCC GCC CGC CGC CCG CCG CCC GCC CAC CGA TAA GCT TGG TCA TGC CCG CCG CCG CCC GCC GCC CGC CGG CCC G{\color{red}T}}} \\
86                          & \scriptsize{\texttt{{\color{red}T}CG CGC CCG GCC GCC CGC CGC CCG CCG CCC GCC CAC CGA TAA GCT TGG TCA TGC CCG CCG CCG CCC GCC GCC CGC CGG CCC GCG C{\color{red}T}}} \\
90                          & \scriptsize{\texttt{{\color{red}T}GC CGC GCC CGG CCG CCC GCC GCC CGC CGC CCG CCC ACC GAT AAG CTT GGT CAT GCC CGC CGC CGC CCG CCG CCC GCC GGC CCG CGC CG{\color{red}T}}} \\
\bottomrule
\end{tabular}
\caption[]
{
Loop sequences for the molecular vice system. Terminal loop thymines, used to compute $R_\textnormal{ee}^\textnormal{hairpin}$ are highlighted (\textcaption{red}).
\newline
\superscript{*} Mismatch locations as used by Fields \etal \cite{Sfields_euler_2013} are highlighted (\textcaption{blue}). Following the experimental design, the mismatch is introduced by changing the sequence of the complement strand.
\newline
\superscript{$\ddagger$} Sequences used by Fields \etal  \cite{Sfields_euler_2013}.
}
\label{tab:cohen-methods-sequences}
\end{table*}

\subsection{Operational definitions of kinking}
\label{sec:cohen-methods-detectors}
\begin{figure}[t]
\includegraphics[width=8.5cm]{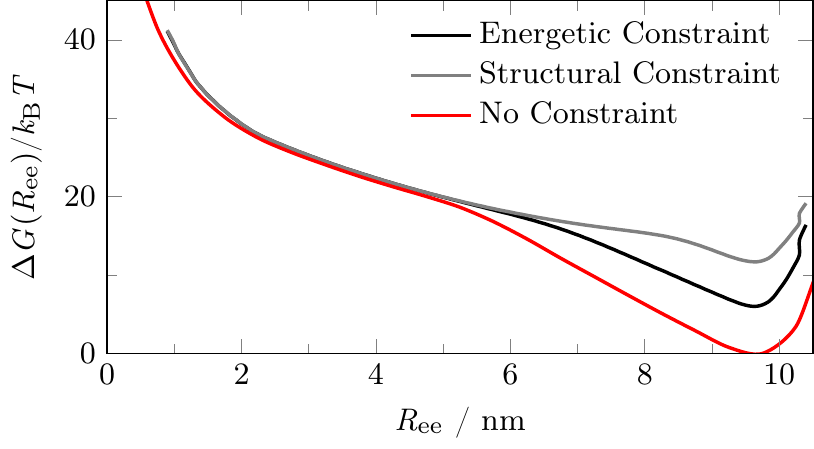}
\caption[]
{
Free energies as a function of $R_\textnormal{ee}$ for a {30}-{bp} duplex constrained to kink by either the energetic or structural criteria. The free energy is normalized by manually matching to the data reported in the main text (\Cref{fig:umbrella-free-energy-bending-D30}\,(a)) for kinked states at small $R_\textnormal{ee}$. Error bars are comparable to line thickness. {Also shown for comparison is the data for a system not constrained to show kinking.}
}
\label{fig:umbrella-kink-high-Ree}
\end{figure}

{
In the main text, we used two `kink detectors' to identify the emergence of kinked states: one with an energetic criterion and one with a structural criterion. We saw that both measures located the transition to a kink-dominated ensemble at approximately the same $R_\textnormal{ee}$. To understand the properties of these detectors further, we performed additional simulations for a {30}-{bp} duplex in which a second umbrella bias was used to forbid unkinked states (separate simulations were performed for each kink criterion). In \Cref{fig:umbrella-kink-high-Ree}, we report $\Delta G(R_\textnormal{ee})$ (the free energy of the kinked state) versus $R_\textnormal{ee}$ obtained from these simulations. Curves were normalized by manually matching to the data obtained for small $R_\textnormal{ee}$ in the original simulations in which unkinked states were permitted -- the zero of free energy is then the same as in \Cref{M-fig:umbrella-free-energy-bending-D30}\,(a).

Both detectors identify deviations from the B-DNA structure of the duplex that are extremely rare in the ensemble of typical states, as can be seen from \Cref{fig:umbrella-kink-high-Ree}. In snapshots of simulations around $R_\textnormal{ee}^\textnormal{trans}$, they also appear to do a good job of distinguishing between clearly kinked and clearly unkinked structures. They are not, however, perfect reporters for configurations that would be intuitively interpreted as `kinks'. Firstly, there is always a degree of arbitrariness involved in defining a kink: is a disruption to the duplex structure of the type that allows enhanced bending a kink even if the net bending angle around the disruption is less than \SI{90}{\degree}, or less than \SI{45}{\degree}? Of our two detectors, the structural criterion is more exacting in this regard, which explains why it reports slightly fewer kinks at $R_\textnormal{ee} \approx R_\textnormal{ee}^\textnormal{trans}$ than the energetic criterion. Although the two detectors do not agree exactly, however, they are reasonably consistent at $R_\textnormal{ee} \lesssim R_\textnormal{ee}^\textnormal{trans}$. This is because duplex disruptions are only common at $R_\textnormal{ee} \lesssim R_\textnormal{ee}^\textnormal{trans}$ {\it because} they allow large bending. We do not observe a significant number of configurations that meet the energetic criterion without large, localized bending because such states do not help to relax the stress applied, and hence remain rare. Thus most configurations that meet the energetic criterion meet the structural criterion (although a few are not quite bent enough). At $R_\textnormal{ee} > R_\textnormal{ee}^\textnormal{trans}$, the number of structures that satisfy the energetic criterion but not the structural criterion is larger, partly because less bending must be localized at kinks to relax the system elsewhwere. Note that for the nicked system, kinking occurs at much larger $R_\textnormal{ee}$ due to the lower cost; hence systems can relax the bending stress without meeting the structural criterion, leading to a bigger discrepancy between the two detectors at $R_\textnormal{ee} \approx R_\textnormal{ee}^\textnormal{trans}$ in this case.

There is a second reason why the curves for the energetic and structural detectors diverge at $R_\textnormal{ee} > R_\textnormal{ee}^\textnormal{trans}$, which is related to a second reason why the detectors are not perfect reporters for kinking. As can be seen in \Cref{fig:umbrella-kink-high-Ree}, the free energy curves for systems constrained by the energetic and structural detectors exhibit a noticeable change in slope as $R_\textnormal{ee}$ gets larger. This change in slope is analogous to the change in slope of the free energy of the unconstrained system, also shown in \Cref{fig:umbrella-kink-high-Ree}. Such a slope change is characteristic of a crossover from one typical conformation (with a certain flexibility) to a qualitatively different one with a distinct flexibility. For our unconstrained system, the slope change is simply the transition between kinked and unkinked states. Thus at $R_\textnormal{ee} \sim 10$\,nm, the typical states that satisfy the energetic and structural criteria for kinking are qualitatively distinct from those that are common at $R_\textnormal{ee} \lesssim R_\textnormal{ee}^\textnormal{trans}$. For example, the energetic criterion can be met by stretching a base pair step, and the structural criterion can be met by flipping out a single base from the helix and rotating it; configurations of this kind are more probable than those that we would intuitively describe as kinked {\it if the duplex is not sharply bent}. These configurations highlight an additional difficulty in unambiguously defining a metric and cost for kinking. However, such states are only common relative to other configurations that satisfy the kink criteria at large $R_\textnormal{ee}$; they are still rare compared to normal states at large $R_\textnormal{ee}$ and remain rare at all values of $R_\textnormal{ee}$ as they do not help to alleviate bending stress. 

In summary, despite the ambiguities in defining and identifying kinks, only duplex disruptions associated with localized bending through large angles are common at $R_\textnormal{ee} \lesssim R_\textnormal{ee}^\textnormal{trans}$ for an intact duplex. Thus the two kink detectors are sufficiently consistent and provide meaningful results.
}

\section{Additional analysis of DNA bending}

\subsection{Bubble size distribution}
\label{sec:umbrella-nhist-bubble-D30}
\begin{figure}
\includegraphics[width=8.5cm]{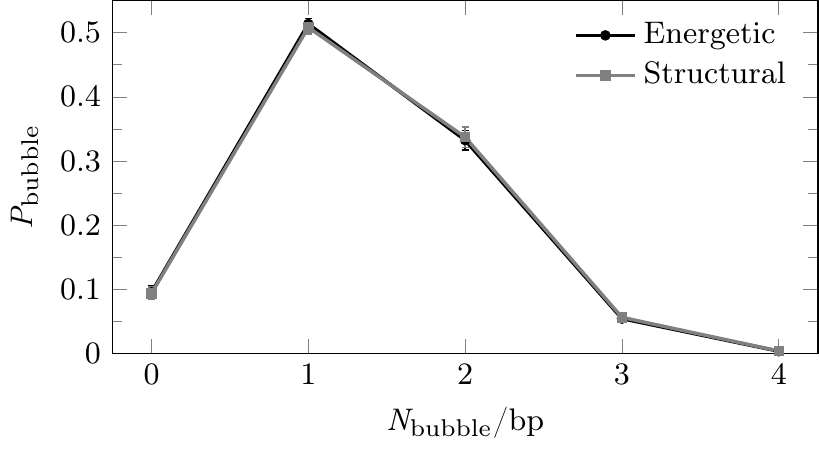}
\caption[]
{
Normalized histogram for the probability distribution ($P_\textnormal{bubble}$) of the total bubble size in base pairs ($N_\textnormal{bubble}$) given the presence of a kink via either the energetic (\textcaption{black}) or structural (\textcaption{grey}) criteria for $N_\textnormal{bp}=\SI{30}{\bp}$ at $\langle R_\textnormal{ee} \rangle \approx \SI{5.3}{\nano\meter}$ from the umbrella potential bin $R_\textnormal{ee}^\prime = \SIrange{4.26}{4.69}{\nano\meter}$. Error bars are of the order of the point size. 
}
\label{fig:umbrella-nhist-bubble-D30}
\end{figure}

The distribution of bubble sizes reveals the frequent presence of a \SIrange{1}{3}{bp} bubble in kinked duplexes (\Cref{fig:umbrella-nhist-bubble-D30}). Briefly, bubbles are defined as loss of base pairs away from the duplex ends, while loss of base pairs at duplex ends is classified as fraying. 
Unsurprisingly, bubbles are almost always co-localized with kinks, so we take total bubble size along the duplex as a proxy for bubbles at a kink. Results for total bubble size given the energetic and structural criteria for a kink are very similar.

\begin{figure}
\includegraphics[width=8.5cm]{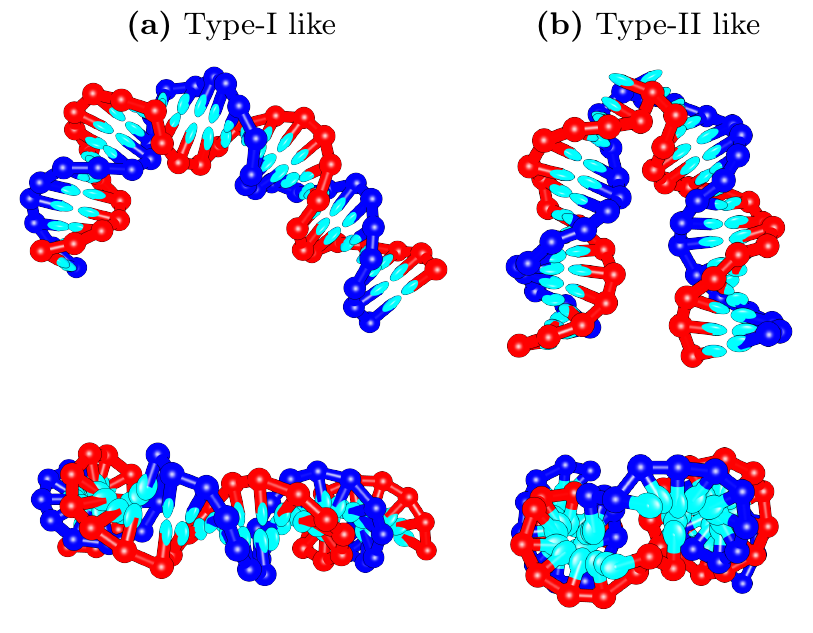}
\caption[]
{
OxDNA representations of type-I like and type-II like kinks in {30}-{bp} duplex. Two perpendicular views are given in each case.
\textbf{(a)} Type-I like kinks have disrupted stacking with preserved base pairing. 
\textbf{(b)} Type-II like kinks disrupt both stacking and base pairing. 
}
\label{fig:umbrella-diagram-kink-typeI-typeII}
\end{figure}

Nomenclature has been developed in all-atom MD simulations \cite{lankas_kinking_2006, Smitchell_atomistic_2011, spiriti_dna_2012} to distinguish between type-I (disruption of stacking but not base pairing) and type-II kinks (disruption of both base pairing and stacking). Due to its coarse-grained nature, we have less confidence that oxDNA has sufficient molecular detail to contribute reliably to this discussion; furthermore, the distinction between type-I and type-II kinks are of less direct relevance to this study. We therefore do not focus on the classification of kink subtypes, although we note that we see configurations which could be assigned to both types (\Cref{fig:umbrella-diagram-kink-typeI-typeII}) and that most kinks involve broken base pairs (i.e.\ are of type-II) -- for example, 
only ~10\% of the kinks for the states contributing to \Cref{fig:umbrella-nhist-bubble-D30} only involve disruption of stacking.

\subsection{Localization of kinks in bending simulations}
\label{sec:D30-umbrella-nhist-kink-location}
\begin{figure}[h!]
\includegraphics[width=8.5cm]{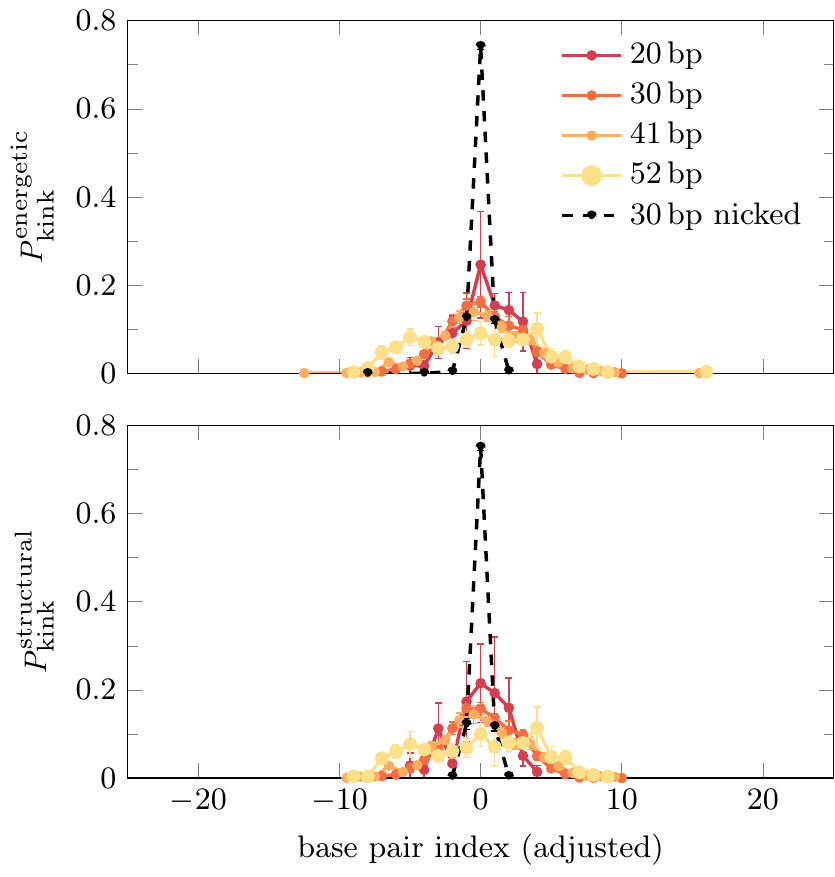}
\caption[]
{
Normalized histogram for the probability distribution of kink localization for the energetic and structural criteria for intact ($N_\textnormal{bp}=20,30,41,52$) and nicked ($N_\textnormal{bp}=30$) duplexes. We sample all duplexes at constant end-to-end distance from the umbrella potential bin $R_\textnormal{ee}^\prime = \SIrange{4.26}{4.69}{\nano\meter}$.
}
\label{fig:umbrella-nhist-kink-location}
\end{figure}

When the ends of a duplex are pulled together, it would be expected that kinking occurs near the midpoint of the duplex. Kinks far from the midpoint would result in incommensurate relaxed duplex sections, which are then less likely to have their ends in close proximity. Analysis of the kink probability distribution confirms that kinks are localized near the midpoint (\Cref{fig:umbrella-nhist-kink-location}). 

Since structural and energetic kinks are a property defined between pairs of consecutive bases (see Definitions section of main text), we report the kink location distribution in base steps, where base step $i$ refers to a kink between base index $i$ and $i+1$. For every sampled state, all base steps along one strand that satisfy the relevant kink criterion are included in the histogram. To normalise, we divide by the total number of detected kinks along both strands in all sampled states. Kinks are considered distinct if they are separated by more than $6$ base steps. {Thus the normalized kink distribution 
gives the probability that a given pair of neighbours contributes to an arbitrarily chosen kink.} To compare multiple lengths, we report the probability of kinking versus the \emph{adjusted} base pair index, centred around the mid-point of the duplex, $(N_\textnormal{bp}-2)/2$. The $-2$ offset accounts for zero-based numbering and indexing the base step to base index $i$ instead of $i+1$.

For the intact duplexes, the distribution is quite wide, $\sim \SI{10}{}$ base pairs, suggesting that entropic effects are quite prevalent and some degree of incommensurately can be tolerated. The distribution also appears to become slightly broader for longer duplexes. Importantly, the preference for localisation about the midpoint of the duplex is because {this is the position that provides the maximum decrease in end-to-end distance}: in relaxed duplex DNA, or in DNA minicircles, one would not necessarily expect a preferred site for kinking with the average-base parameterisation of oxDNA. 

For the nicked duplex at $N_\textnormal{bp}=\SI{30}{\bp}$, the distribution is quite narrow, $\sim \SI{2}{}$ base pairs, reflecting the strong preference for localisation of kinking at the nick (\Cref{fig:umbrella-nhist-kink-location}). Contrary to the intact duplexes, the preference for kinking at a nick is not just due to the bending stress imposed on the system: nicked sites have a lower free energy cost for kinking because base pairing need not be disrupted for a kink to form; {only stacking must be compromised}.

\subsection{Re-entrant continuous bending}
\label{sec:umbrella-diagram-small-ell-continious-bending}

\begin{figure}[t]
\includegraphics[width=8.5cm]{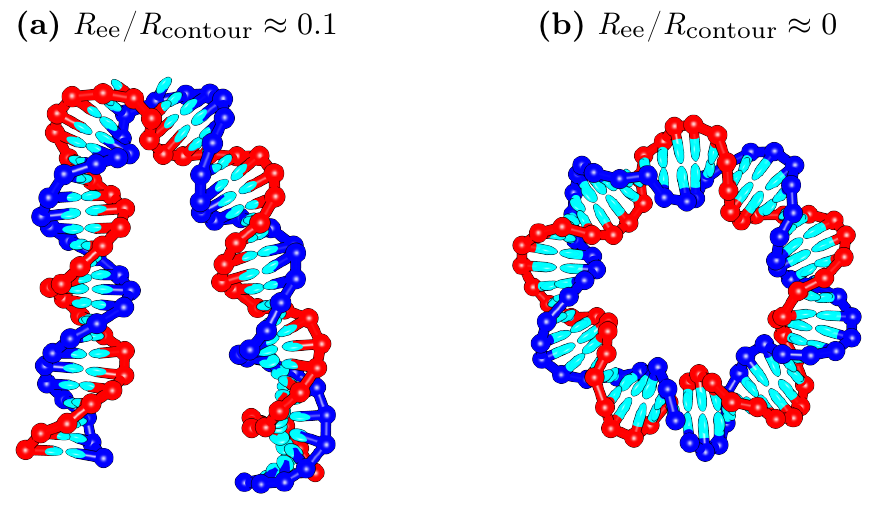}
\caption[]
{
Representative OxDNA configurations for a {52}-{bp} duplex at small $R_\textnormal{ee}/R_\textnormal{contour}$.
{\textbf{(a)} A kinked state, with $R_\textnormal{ee}$ small but greater than 2\,nm. 
\textbf{(b)} An unkinked, minicircle-like state with $R_\textnormal{ee} \ll 2$\,nm. With this constraint, a single kink cannot relax the majority of the bending stress, unlike in case (a).
}}
\label{fig:umbrella-diagram-small-ell}
\end{figure}

At $R_\textnormal{ee}/R_\textnormal{contour} \approx 0$, kinking appears to be
less effective at relieving bending stress; for all but the shortest duplexes,
there is re-entrant behaviour from kinking back to continuous bending. A
typical kinked structure is shown in
\Cref{fig:umbrella-diagram-small-ell}\,(a); it consists of two long (relatively
relaxed) duplex sections and a highly bent kink region. {Given that
$R_\textnormal{ee}$ is measured in terms of the distance between terminal base
pair centres}, however, such a configuration cannot bring the ends into
extremely close proximity ($R_\textnormal{ee} < \SI{2}{\nano\meter}$, {the
approximate duplex diameter}) due to steric effects. 
Instead, the blunt ends of the duplex must face each other to achieve 
$R_\textnormal{ee} < \SI{2}{\nano\meter}$ and this 
fundamentally changes the constraints placed on the rest of the system, and
makes kinking opposite the blunt ends less effective at relieving the stress.

The DNA duplex adopts minicircle-like configurations for extremely small values of $R_\textnormal{ee}$, as shown in \Cref{fig:umbrella-diagram-small-ell}\,(b). Minicircles can, of course, undergo kinking; however, the geometrical constraints mean that it is impossible to localise the majority of bending stress into a single kink \cite{Syan_localized_2004}, unlike in systems with a slightly larger separation of the blunt ends. Thus we obtain the seemingly paradoxical result that kinking is more likely for systems that have some flexibility in their ends than for systems, such as minicircles, where the ends are constrained to be in extremely close proximity.

\subsection{Kink transition free energy}
For clarity, we present the bending free energy and end-to-end separation at the mid-point of the kink transition for all simulations in \Cref{tab:free-energy-kink-formation}.

\begin{table}[h!]
\ra{1.3}
\begin{tabular*}{.4\textwidth}{@{\extracolsep{\fill}}rrr}
\toprule
\multicolumn{1}{l}{} &  \multicolumn{1}{r}{$\Delta G_\textnormal{trans}$ / \SI{}{\kT}} & \multicolumn{1}{r}{$R_\textnormal{ee}^\textnormal{trans}$ / \SI{}{\nano\meter}} \\
\midrule
\multicolumn{3}{l}{Length / bp} \\ \cmidrule(l){1-3}
20                      & 15.9   & 4.7   \\
30                      & 18.1   & 5.5   \\
\superscript{n}30       &  6.0   & 8.0   \\
41                      & 19.7   & 5.3   \\
52                      & 21.4   & 4.3   \\ \\
\multicolumn{3}{l}{Sequence / \SI{28}{bp}} \\ \cmidrule(l){1-3}
poly(GC)      & 21.1   & 5.1  \\
Avg-base      & 18.0   & 5.4  \\
poly(AT)      & 14.8   & 5.8  \\ \\
\multicolumn{3}{l}{Mismatch / \SI{28}{bp}} \\ \cmidrule(l){1-3}
Intact                     & 16.1   & 5.7  \\
\superscript{m}GG          & 13.0   & 6.0  \\
\superscript{m}GT          & 13.3   & 6.0  \\
\superscript{m}AA          & 12.8   & 6.1  \\
\superscript{m}CC          & 13.5   & 5.9  \\ \\
\multicolumn{3}{l}{Nicked / \SI{28}{bp}} \\ \cmidrule(l){1-3}
\superscript{n}poly(GC)      &  7.9   & 7.2  \\
\superscript{n}Avg-base      &  6.7   & 7.3  \\
\superscript{n}poly(AT)      &  4.8   & 7.6  \\\\
\bottomrule
\end{tabular*}
\caption[]
{
{The bending free-energy $\Delta G_\textnormal{trans}$ and the end-to-end distance $R_\textnormal{trans}$ at the midpoint of the kink transition, which we define as the point at which kinks are present in \SI{50}{\percent} of states according to the energetic detector, for all the different systems studied here. Results for 62 and 73\,bp are not given because although the kinked microstates occur, the probability of kinking never reaches 0.5.}
\newline
\superscript{n} Nick at the mid-point of the duplex.
\newline
\superscript{m} Single mismatch in the duplex with the same sequences
as used by Fields \etal \cite{Sfields_euler_2013}.
}
\label{tab:free-energy-kink-formation}
\end{table}

\FloatBarrier
\newpage
\section{Additional data for the molecular vice}
\subsection{Long-length limit}
\label{sec:cohen-Lall-long}
In the main text, we show the structural properties for variable loop length ($N_\textnormal{loop}$) and duplex length ($N_\textnormal{dup}$) (\Cref{M-fig:cohen-Lall}). {The data suggest that kinked configurations are gradually superseded by continuously bent states as $N_\textnormal{loop}$ increases}. {This behaviour is also} observed in our investigation of length-dependence in the free-energy of bending of intact duplexes (\Cref{M-fig:umbrella-free-energy-bending-length}), and would therefore be expected for the molecular vice.

\begin{figure}[t]
\includegraphics[width=8.5cm]{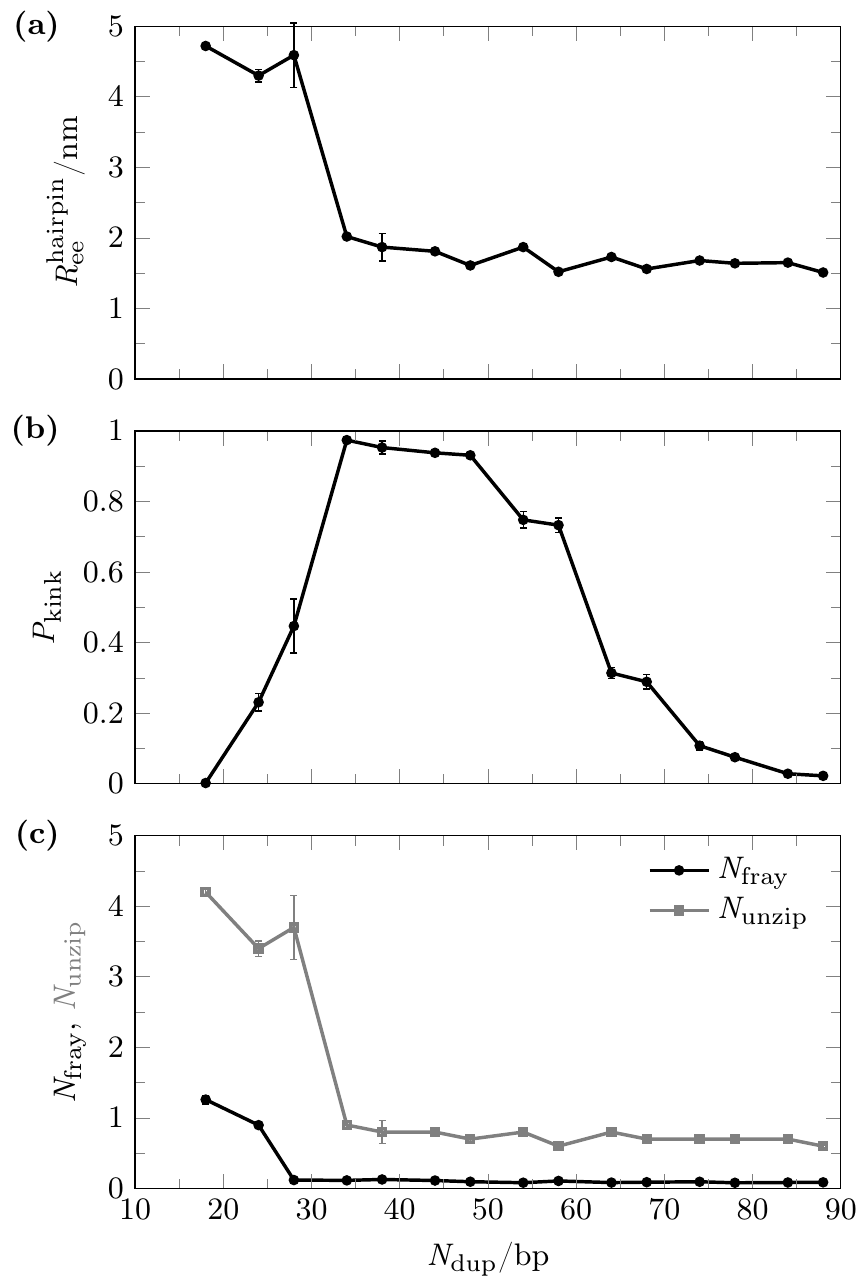}
\caption[]
{
Length-dependence of \textbf{(a)} hairpin separation $R_\textnormal{ee}^\textnormal{hairpin}$, \textbf{(b)} probability of kinking $P_\textnormal{kink}$ and \textbf{(c)} fraying ($N_\textnormal{fray}$) and stem unzipping ($N_\textnormal{unzip}$) for the molecular vice in the $N_\textnormal{dup} = (N_\textnormal{loop}-2)$ 
limit.
}
\label{fig:cohen-Lall-long}
\end{figure}

{We make the length-dependence of kinking, hairpin separation, fraying and stem unzipping in the molecular vice more explicit in \Cref{fig:cohen-Lall-long}, by profiling the longest $N_\textnormal{dup}$ at a range of $N_\textnormal{loop}$ lengths.} {At the lowest values of $N_\textnormal{loop}$, unzipping (and a small amount of fraying) is the dominant method of relieving stress. At intermediate $N_\textnormal{loop}$, kinked states dominate.} A transition from kinked to continuously bent states is observed at long $N_\textnormal{dup}$ (and therefore long $N_\textnormal{loop}$), with the midpoint at $N_\textnormal{dup} = \SIrange{58}{64}{bp}$. By $N_\textnormal{dup}={88}$\,bp, duplexes are almost exclusively continuously bent ($P_\textnormal{kink} < \SI{2}{\percent}$). Unsurprisingly, the length associated with this crossover is consistent with our bending simulations (\Cref{M-fig:umbrella-free-energy-bending-length}).

\begin{figure}[t]
\includegraphics[width=8.5cm]{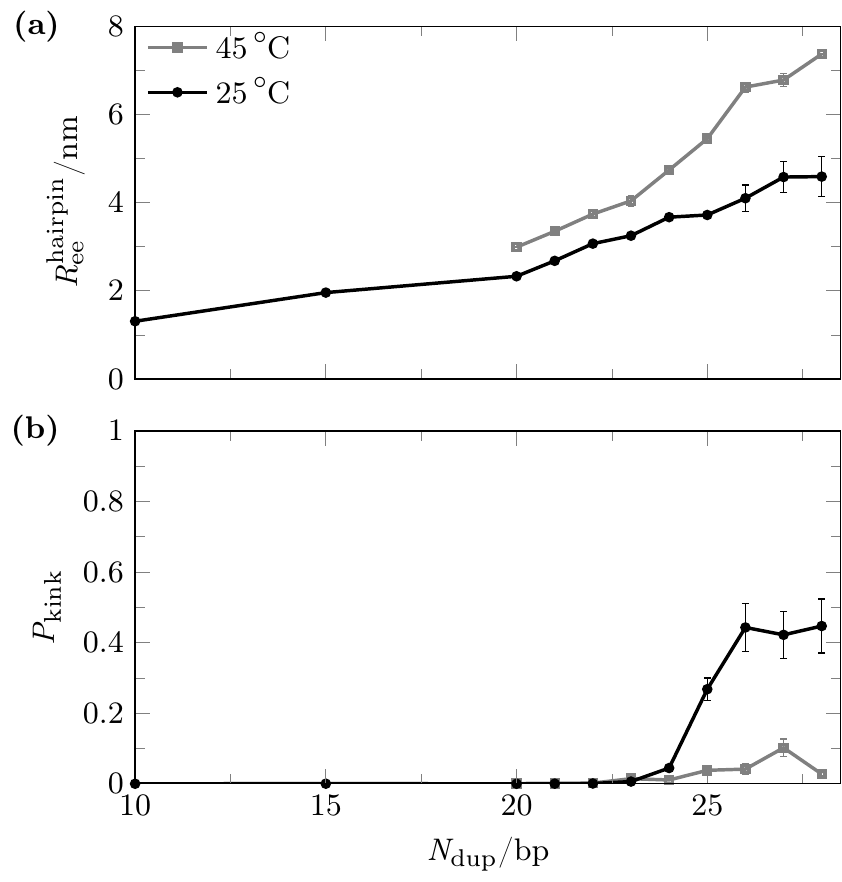}
\caption[]
{
Temperature-dependent structural properties of the molecular vice for $N_\textnormal{loop}=30$ at $T=\SI{25}{\celsius}$ (\textcaption{black circle}) and $T=\SI{45}{\celsius}$ (\textcaption{gray square}). 
}
\label{fig:cohen-temperature-L30}
\end{figure}

\subsection{Temperature-dependence}
\label{sec:cohen-temperature-L30}
We cannot currently study the exact salt concentration used by Fields \etal \cite{Sfields_euler_2013} with oxDNA. However, we can probe the impact of increasing temperature, which may also be somewhat analogous to decreasing salt concentration, as both serve to weaken base pairing; in particular, reducing the unzipping force associated with the hairpin \cite{huguet_single-molecule_2010}. We also note that temperature is not perfectly controlled in gel-based experiments due to electrophoresis-induced heating of the gel running buffer. 
    
Temperature-dependence was explored with additional simulations at \SI{35}{\celsius} (data not shown) and \SI{45}{\celsius} (\Cref{fig:cohen-temperature-L30}). We find negligible kinking at $T=\SI{45}{\celsius}$ for $N_\textnormal{loop}=30$, consistent with the reasoning that weakened base pairing in the stem leads to easier unzipping.

\end{document}